\documentstyle[preprint,tighten,aps]{revtex}

\newcommand{\ewxy}[2]{\setlength{\epsfxsize}{#2}\epsfbox[30 60 540 570]{#1}}

\newcommand{\delfour}{{\Delta^{(4)}}}
\newcommand{\delsq}{\Delta^{(2)}}

\newcommand{\vev}[1]{\langle #1 \rangle}
\newcommand{\Mbz}{{(a_sM_0)}}

\newcommand{\Bv}{{\bf \tilde{B}}}

\newcommand{\sigmav}{\mbox{\boldmath$\sigma$}}

\newcommand{\act}{{\cal S}}
\newcommand{\be}{\begin{equation}}
\newcommand{\ee}{\end{equation}}
\newcommand{\nl}{\nonumber \\}

\begin{document}
\preprint{\begin{minipage}{2in}\begin{flushright}
 OHSTPY-HEP-T-02-008 \\ GUPTA/02/07/01 \end{flushright}
  \end{minipage}}
\input epsf

\title{ {\bf Semileptonic $B$ Decays from an NRQCD/D234 Action}}

\author{J.Shigemitsu$^a$,
  S.Collins$^{b}$, C.T.H.Davies$^b$,
 J.Hein$^{c}$, \\
  R.R.Horgan$^{d}$,
G.P.Lepage$^e$.\\[.4cm]
\small $^a$Physics Department, The Ohio State University,
 Columbus, OH 43210, USA. \\[.2cm]
\small $^b$Department of Physics \& Astronomy, University of Glasgow, 
 Glasgow, G12 8QQ, UK. \\[.2cm]
\small $^c$Department of Physics \& Astronomy, University of Edinburgh, 
 Edinburgh, EH9 3JZ, UK. \\[.2cm]
\small $^d$D.A.M.T.P. , CMS, Wilberforce Road, Cambridge, England CB3 0WA, 
UK. \\[.2cm]
\small $^e$Newman Laboratory of Nuclear Studies, Cornell University, 
 Ithaca, NY 14853, USA. 
\\ }


\maketitle

\begin{abstract}
\noindent
Semileptonic, $B \rightarrow \pi \; l \overline{\nu}$, decays are 
studied on quenched anisotropic lattices using tree-level tadpole 
improved Symanzik glue, NRQCD heavy quark and D234 light quark
actions.  Constrained fitting methods are applied to extract 
 groundstate contributions to two-point and three-point correlators.
 We agree with previous lattice determinations of the form factors.  The 
major source of systematic error here, as in previous work, comes from 
the chiral extrapolation to the physical pion mass. Future calculations 
must work at lighter quark masses to resolve this.

\end{abstract}

\vspace{.2in}
\noindent
PACS numbers: 12.38.Gc, 13.20.He, 14.40.Nd

\newpage

\section{Introduction}

\noindent
Determinations of the CKM matrix element $|V_{ub}|$ from exclusive 
semileptonic $B$ decays rely on lattice input for the 
$\langle \pi | J_\mu|B\rangle$ or 
$\langle \rho | J_\mu|B\rangle$ matrix elements.  Reducing errors in  
these lattice calculations and extending their kinematic range will be an 
important contribution to consistency tests of the CKM matrix, and thereby 
of the Standard Model.  Several quenched studies of $B \rightarrow 
\pi \; l \overline{\nu}$ decays have been carried out in recent years 
employing a wide range of different actions and methods 
\cite{fnal,jlqcd,ape,ukqcd}. 
 Despite 
the differences in sources of systematic errors there is general agreement in 
final results, especially for the form factor $f_+(q^2)$ which is 
directly relevant for the differential decay rate measured in experiments.

\vspace{.1in}
\noindent
The difficulties with lattice simulations for these semileptonic
form factors are well known.  Lattice calculations are most reliable 
at large $q^2 \equiv (p_\pi-p_B)^2$, i.e. close to the zero recoil
point, whereas experiments are limited to smaller $q^2$.
In the $B$ meson rest frame, small $q^2$ implies large pion momenta 
which introduces both $ap_\pi$ discretization errors on the lattice and 
large statistical errors.  Furthermore it is becoming increasingly 
more evident that a major source of systematic error comes from 
the chiral extrapolation to a physical final state pion.

\vspace{.1in}
\noindent
This article describes further investigations of $B \rightarrow \pi \; 
l \overline{\nu}$ decays in the quenched approximation, where 
we have experimented with ways to improve errors due to finite pion 
momenta.  The first difference from previous work is the use of more 
highly improved actions.  We employ Symanzik glue rather than 
the Wilson glue action, the D234 light quark action rather than clover and 
 an NRQCD action corrected through $O(a^2)$.  We find good dispersion 
relations for pion energies up to at least $1.5$GeV.  The ability or 
inability to cover a $q^2$ range  overlapping with experiment is 
determined more by the other challenges, namely statistical noise and chiral 
extrapolations, than by discretization errors.  This will be even more 
true if one goes to lattices finer than the relatively coarse, 
$1/a_s = 1.2$GeV, lattices used in the present study. 

\vspace{.1in}
\noindent
The other two new ingredients here are the use of an anisotropic lattice and 
constrained fitting methods. As demonstrated in many contexts, anisotropic
 lattices lead to correlators that have a larger number of
 data points inside the time 
region where signal-to-noise is still good. 
 This allows for more accurate 
extraction of energies and amplitudes \cite{mikecolin,manke,ianron,comparison}.
 Reference \cite{comparison} showed the advantages 
of anisotropic lattices for correlators involving finite momentum hadrons.
In the present work we add one more major improvement to this scheme, 
 namely constrained fits \cite{bayes}.  These 
 new analysis tools, based on Bayesian statistics,
 allow us to increase the number 
of exponentials in fits to single correlators, without losing stability 
or having errors in low lying energies become large.  One typically fits 
to all (or almost all) data points, with $t_{min}=0$ or $=1$, 
irrespective of where or whether 
a plateau sets in.  Hence one can take full advantage 
of all the data points with small statistical errors that the anisotropic 
lattice gives us.  Previously if one just went to anisotropic lattices 
and used conventional fitting methods one had to rely on having 
excellent smearings or a large matrix of smearings 
to ensure overlap between a plateau region and 
the region with good signal-to-noise.

\vspace{.1in}
\noindent
We find that constrained fits allow us to extract groundstate contributions 
to two-point and three-point correlators in a controlled way despite the fact
 that our smearings are far from optimal.  
Nevertheless, 
our final results for $B \rightarrow \pi \; l \overline{\nu}$ form factors
still exhibit large systematic errors, comparable to those quoted by other 
groups.  The main reason for this is that our systematic error is 
dominated by the chiral extrapolation uncertainties.  The D234 light quark 
action suffers, in common with the clover and Wilson actions, from 
exceptional configurations. 
This limits our ability to go 
to small pion masses and necessitates a large chiral extrapolation 
(we work in the range $0.7 m_{strange} < m_q < 1.3 m_{strange}$).
Even the correct ansatz for the extrapolation is unclear at the present
 time.
In the future, rather than try to go to finer lattices with the same action,
we believe that, in order to improve on chiral extrapolation errors,
 one needs to go to light quark actions with good chiral 
properties. Work with improved staggered light quarks, for instance,
 has already started \cite{staggered,matt}.\\
Difficulties with chiral extrapolations also limit our ability to go
to pion energies larger than $1$GeV (or $q^2 < 16$GeV$^2$).  If one were to 
work harder and double the statistics one might be able to consider pions 
with momenta up to $1.5$GeV and 
 go down to $q^2 \sim 12$Gev$^2$ as long as the light quark mass 
stays around the strange 
mass.  Nevertheless, carrying out chiral extrapolations at large 
pion momenta would 
introduce prohibitively large errors. Recently the idea has been put 
forward of working in a reference frame where the entire $q^2$ range 
can be covered with pions having momenta less than $1$GeV  (see ref. 
\cite{mnrqcd} for 
discussion of the ``Moving NRQCD'' formalism).  We believe the 
future of accurate semileptonic $B$ decay studies lies in a combination 
of using light quark actions with good chiral properties and a formalism 
that can handle 
 $B$ mesons decaying at large velocities.  The lessons 
derived from the present project should be very useful 
in such future work, especially in 
the analysis of three-point correlators.

\vspace{.1in}
\noindent
In the next section we introduce the actions used in this project.  Section 
III provides simulation details and discusses 
results from two-point correlators.
Section IV describes our direct fits to three-point correlators.  This step
replaces the conventional approach of considering ratios of three- 
and two-point correlators and looking for a plateau which is then
 fit to a constant.  In section V we discuss chiral extrapolations
and  present results for form factors. Comparisons are made with previous 
lattice work. 
 We then conclude with a summary section.
We delegate details of constrained fitting methods to an appendix.  A further 
appendix lists heavy-light current one-loop matching coefficients for 
the action used in this article.

\section{Gauge and Quark Actions }

\noindent
The gauge and quark actions of this article are the anisotropic actions 
discussed in ref. \cite{comparison}. 
 For the glue we use the tree-level tadpole-improved 
Symanzik action  with rectangles only in spatial directions.
\begin{eqnarray}
\act_G &=& - \beta
 \sum_{x,\,s > s^\prime} \frac{1}{\chi_0} \left\{
\frac{5}{3} \frac{P_{ss^\prime}}{u_s^4} 
- \frac{1}{12} \frac{R_{ss^\prime}}{u_s^6} 
- \frac{1}{12} \frac{R_{s^\prime s}}{u_s^6} \right\} \nl
 & & - \beta \;\; \sum_{x,s} \chi_0 \left\{
\frac{4}{3} \frac{P_{st}}{u_s^2 u_t^2} 
- \frac{1}{12} \frac{R_{st}}{u_s^4u_t^2} \right\} .
\label{gaction}
\end{eqnarray}
$P_{\mu \nu}$ and $R_{\mu \nu}$ denote plaquettes and rectangles 
in the $\mu \nu$ plane. 
The variables $s$ and $s^\prime$ run only over spatial directions and 
 $u_t$ and $u_s$ are the tadpole-improvement ``$u_0$'' factors for temporal 
and spatial link variables respectively.  We use the Landau link
definition of $u_0$ in this article.
$\chi_0$ is the bare anisotropy which differs from the true or renormalized 
anisotropy,
\be
\chi \equiv a_s/a_t,
\ee
in the presence of quantum corrections.  
We use torelon dispersion relations \cite{markron}
 to fix $\chi$ nonperturbatively.  
Starting with $\chi_0=3.$ at $\beta=2.4$ we find the renormalized 
anisotropy to be $\chi=2.71(3)$.

\vspace{.1in}
\noindent
The light quark action is the D234 action of ref.\cite{alford}
 modified for anisotropic 
lattices.
\begin{eqnarray}
\label{sd234c}
\act^{(aniso)}_{D234} &=& a_s^3 a_t \sum_x \overline{\Psi}_c 
\left\{  \gamma_t \frac{1}{a_t} \nabla_t 
 + \frac{C_0}{a_s} \vec{\gamma} \cdot (\vec{\nabla} -
 \frac{1}{6} 
C_{3} \vec\nabla^{(3)}) + m_0 \right. \nl
 &  & - \frac{r a_s}{2} \left [ \frac{1}{a_t^2}  
 \nabla_t^{(2)} 
+  \frac{1}{a_s^2} \sum_{j=1}^3 ( \nabla_j^{(2)} 
- \frac{1}{12} C_{4} \nabla_j^{(4)} ) \right ] 
 - r a_s \left. \frac{C_F}{4} 
\frac{i \sigma_{\mu \nu} \tilde{F}^{\mu \nu}} {a_\mu a_\nu} 
\right\} 
\Psi_c    \\
\label{lqaction}
 &=&  \sum_x \overline{\Psi}
\left\{  \gamma_t \nabla_t +
\frac{C_0}{\chi} \vec{\gamma} \cdot (\vec{\nabla} -
 \frac{1}{6} 
C_{3} \vec\nabla^{(3)}) + a_t m_0 \right. \nl
 &  & - \frac{r }{2} \left [ \chi  \nabla_t^{(2)} 
+  \frac{1}{\chi} \sum_{j=1}^3 ( \nabla_j^{(2)} 
- \frac{1}{12} C_{4} \nabla_j^{(4)} ) \right ] 
 - r \left. \frac{C_F}{4} 
i \sigma_{\mu \nu} \tilde{F}^{\mu \nu} \frac{a_s a_t}{a_\mu a_\nu} 
\right\} 
\Psi 
\end{eqnarray}
The quark fields $\Psi_c$ and the dimensionless lattice 
fields $\Psi$ are related through
\be
\label{psiscale}
\Psi = a_s^{3/2}  \Psi_c .
\ee
Definitions of lattice derivatives and the improved $\tilde{F}_{\mu \nu}$ 
are given, for instance, in the appendix to ref.\cite{pert}.
  We work with 
$r$, $C_F$, $C_3$ and $C_4$ all set equal to one.  The ``speed of light'' 
coefficient, $C_0$, is tuned perturbatively or
 nonperturbatively using pion dispersion 
relations.  Its actual value will be discussed in the next section.

\vspace{.1in}
\noindent
For the heavy quark we use the standard NRQCD evolution equations which 
follow from the action \cite{cornell,ianron2},
\begin{eqnarray}
 \label{hqaction}
&& \act_{NRQCD} =  \nl
&& \sum_x \left\{  \overline{\Phi}_t \Phi_t - 
 \overline{\Phi}_t
\left(1 \!-\!\frac{a_t \delta H}{2}\right)_t
 \left(1\!-\!\frac{a_tH_0}{2n}\right)^{n}_t
 U^\dagger_4
 \left(1\!-\!\frac{a_tH_0}{2n}\right)^{n}_{t-1}
\left(1\!-\!\frac{a_t\delta H}{2}\right)_{t-1} \Phi_{t-1} \right\}.
 \end{eqnarray}
 $H_0$ is the nonrelativistic kinetic energy operator,
 \be
a_t H_0 = - {\delsq\over2\chi\Mbz},
 \ee
and $\delta H$ includes $1/M$ relativistic and 
 $O(a^2)$ finite lattice spacing
corrections,
 \begin{eqnarray}
a_t\delta H 
&=& - \frac{1}{2\chi\Mbz}\,\sigmav\cdot\Bv 
  + \frac{\delfour}{24\chi\Mbz}.
\label{deltaH}
\end{eqnarray}
All derivatives are tadpole improved and,
\be
\delsq = \sum_{j=1}^3\nabla_j^{(2)}, \qquad \qquad \delfour = \sum_{j=1}^3
\nabla_j^{(4)}
\ee

\vspace{.1in}
\noindent
The leading discretization errors in the total action are $O(a_s \, 
\alpha_s)$ errors coming from the light quark action.  The leading 
relativistic corrections are $O(\alpha_s \frac{\Lambda_{QCD}}{M})$ 
coming from one-loop corrections to tree-level coefficients in the
 NRQCD action.  We will work with heavy-light currents corrected 
to the same level as the action, i.e. we will include $O(\alpha_s)$ 
and $O(\Lambda_{QCD}/M)$ terms but not $O(a_s \, \alpha_s)$ or 
$O(\alpha_s \frac{\Lambda_{QCD}}{M})$ corrections.  We estimate 
systematic errors from the latter terms to be at the 8\% and 3\% 
levels respectively.

\section{ Simulation Details }

\noindent
Table I summarizes lattice and action parameters. We work on $12^3 \times 48$ 
quenched anisotropic lattices with $\chi \equiv a_s/a_t =2.71(3)$, 
as determined 
from torelon dispersion relations in the pure glue theory.  We use a 
total of 199 configurations and run both time-forward  and time-reversed 
NRQCD evolutions in order to increase statistics.  The former uses
timeslices 0 - 23, and the latter timeslices 0,47 ,46 ....25.
For the $B$ meson correlators and the three-point correlators 
we find little evidence for correlations between the two runs and 
 close to
 a $\sqrt{2}$ improvement in statistics.  
  Nevertheless, in  our data analysis we 
always bin the data from the two time evolutions before carrying out fits.

\vspace{.1in}
\noindent
Based on string tension calculations we estimate the spatial lattice 
spacing to correspond to $1/a_s = 1.20(5)$GeV.  The $\rho$ mass gives 
a similar value of $(1/a_s)_\rho = 1.18(6)$GeV.

\vspace{.1in}
\noindent
We have carried out simulations at five values of the bare light 
quark mass corresponding to $P/V$ = 0.624(13), 0.675(8), 0.714(14),
0.736(6) and 0.760(6).  In terms of $a_t m \equiv a_t (m_0 - m_{crit.})$, 
we have the range $a_t m$ = 0.023, 0.028, 0.033, 0.038 and 0.043. 
The middle value of $a_t m = 0.033$ is very close to the strange quark 
mass as determined by the $\phi$.  We find $M_V = 1.003(20)(40)$GeV 
(second error comes from the uncertainty in scale which we take from the 
string tension) compared 
to the experimental $M_\phi = 1.019$GeV. Hence our light quark masses span 
roughly the range from $0.7 m_s$ to $1.3 m_s$.  Clearly one would ideally 
like to go to much smaller quark masses.  For our lightest mass we encountered
 one exceptional configuration in an ensemble of initially 200 configurations.
 This stopped us from attempting to further decrease the quark mass and 
reduced the number of usable configurations to 199.  
For the heavy quark mass we employed one value
 $a_s M_0 = 4$, tuned
to be close to the $b$ quark mass using the $B_s$ meson mass as experimental 
input.  Fig.1 shows $a_t M_{kin}$ for the $B_s$ meson extracted from 
$B$ correlators with different spatial momenta. 
\be
M_{kin} = \frac{p^2 - \delta E(p)^2}{2 \, \delta E(p)}
\ee
with $\delta E(p) = E_B(p) - E_B(0)$.  $E_B$ denotes the falloff energy 
of $B$ correlators and differs from the total energy of the $B$ meson 
since the NRQCD action does not include a rest mass term.  This distinction
is irrelevant for the difference $E_B(p) - E_B(0)$.  
Data from different momenta all give 
results consistent with the experimental value for $M_{B_s}$ and also with 
perturbative expectations.  The 
latter is based on
\be
\label{mkin}
M_{kin}^{pert} = Z_m M_0 - E_0 + E_B(0).
\ee
The full horizontal line in Fig.1 gives the one-loop result for 
$a_t M_{kin}^{pert}$.  The two dotted lines are estimates of errors due to
higher order corrections. 
In ref.\cite{jlqcd} it was found 
that semileptonic form factors in the $b$ quark region are not very 
sensitive to the heavy 
quark mass.  Hence we believe errors coming from inadequate tuning of 
$a_s M_0$ are small compared to our other systematic errors.

\vspace{.1in}
\noindent
The ``speed of light'' coefficient $C_0$ in the light quark action, 
eq.(\ref{lqaction}), was fixed using pion dispersion relations. 
We found it sufficient to consider the lowest pion momentum at 
one value of the bare mass ($a_t m = 0.033$).  This fixed $C_0 = 0.94$ 
which works well for all 5 light quark masses.  Fig.2 shows 
the speed of light $C(p)$ defined as,
\be
C(p) = \sqrt{\frac{E_\pi^2(p) - E_\pi^2(0)}{p^2}}
\ee
for three different light quark masses.  One sees that the 
relativistic dispersion relation holds well, within errors, for 
momenta up to at least $1.5$GeV.  In our analysis of three-point correlators 
and extraction of form factors we will use continuum relativistic 
formulas for $E_\pi(p)$.  In this work it was fairly painless to fix $C_0$ 
nonperturbatively.  Had we used one-loop perturbation theory \cite{pert} we
would not have been too far off with $C_0^{1-loop} = 0.91(3)$.  Once 
 two-loop results are known, we will probably be able to dispense 
with nonperturbative tunings of parameters such as $C_0$ or the 
renormalized anisotropy $\chi$ and anisotropic actions will be just 
as easy to handle as isotropic ones \cite{hartron}.

\vspace{.1in}
\noindent
For completeness we show chiral extrapolations for $m_\rho$ and 
$E_B(0)$ in Fig.3.  This leads to the $(1/a_s)_\rho $ quoted 
above and to a $B_s - B_d$ splitting of $86(13)$MeV.  The experimental 
$B_s-B_d$ mass difference is $90.2(2.2)$MeV.  

\section{Analysis of Three-point Correlators}

\noindent
The first step in semileptonic $B$ decay simulations on the lattice is to 
calculate the three-point correlator,
\be
\label{threepnt1}
C_\mu^{(3)}(\vec{p}_B,\vec{p}_\pi,t_B,t) = \sum_{\vec{x}}\sum_{\vec{y}}
e^{-i\vec{p}_B \cdot\vec{x}} e^{i(\vec{p}_B - \vec{p}_\pi)\cdot \vec{y} }
\vev{0|\Phi_{B}(t_B,\vec{x})\,V^L_\mu(t,\vec{y})\, \Phi^\dagger_\pi(0)
|0} .
\ee
$\Phi_\pi^\dagger$ and $\Phi_B^\dagger$ are interpolating operators 
used to create the pion or $B$ meson respectively.  $V^L_\mu$ is the 
dimensionless Euclidean space 
lattice heavy-light vector current. 
The continuum Minkowski space $V_\mu$ is related to $V^L_\mu$ via
\be
V_\mu = a_s^{-3} Z_{V_\mu}\,\xi(\mu)\, V^L_\mu .
\ee
$\xi(\mu)$ is the conversion factor between Minkowski and Euclidean 
space $\gamma$-matrices and $Z_{V_\mu}$ is the heavy-light current 
matching coefficient.  Perturbative estimates of $Z_{V_\mu}$ are 
discussed in an appendix.  \\
We work exclusively with $B$ mesons decaying at rest ($\vec{p}_B = 0$).  
$t_B$ is fixed at $t_B=23$ for time-forward runs and $t_B=25$ in 
time-reversed runs.

\vspace{.1in}
\noindent
We have found it convenient to rescale $C_\mu^{(3)}$ as follows,
\be
\label{rescale}
\tilde{C}^{(3)}_\mu (p_\pi,t) = \xi(\mu) \,C^{(3)}_\mu(\vec{p}_B=0,p_\pi,
t_B,t) \, e^{E^{(1)}_B (t_B-t)} 
\ee
where $E_B^{(1)}$ is the ground state $B$ meson falloff energy, obtained 
from fits to two-point correlators (in a bootstrap analysis $E_B^{(1)}$ 
is obtained separately for each bootstrap ensemble).  $\tilde{C}^{(3)}_\mu$ 
is then fit to,
\begin{eqnarray}
\tilde{C}^{(3)}_\mu(p_\pi,t)  &=& 
  \left\{ \sum_j^{N_B} \sum_l^{N_\pi}
 A_{jl}e^{-E_\pi^{(l)}  t} e^{-E_B^{(j)}(t_B-t)}
\right\} e^{E_B^{(1)}(t_B-t)} \\
&=& \sum_l^{N_\pi} A_{1l}e^{-E_\pi^{(l)}t} \nonumber  \\
\label{thrpnt}
&+& \sum_l^{N^\prime_\pi}
 A_{2l}e^{-E_\pi^{(l)}t} e^{-(E_B^{(2)} - E_B^{(1)})(t_B-t)} 
 \\
&+& .......... \nonumber
\end{eqnarray}
We use constrained (Bayesian) fits of ref.\cite{bayes} to fit a single 
correlator to the multi-exponential form on the RHS of eq.(\ref{thrpnt}).
Details are given in the appendix.
Fits are carried out with $t_{min}$ fixed at $t_{min}=1$ and for various 
$t_{max} < t_B$.  The number of exponentials is increased until a 
good fit to the data is obtained.  For a ``good fit'' we generally 
require $\chi^2/d.o.f. \le 1$.  In a few cases we accept $\chi^2/d.o.f $
as high as 
$\sim 1.2$ but even in those instances the $Q$-values are larger than 0.2 .
 Sample fit results for $\langle V^L_0 \rangle$ are  
shown in Figs.4 and 5.  For pion momenta (001), (011) and (111) and using 
$t_{max} = 17 \sim 21$ good fits were obtained with just the first line 
on the RHS of (\ref{thrpnt}) and $N_\pi \ge 2$.  The figures show results 
for $N_\pi=3$ and $t_{max}=19$. 
 For the zero momentum case an additional excited $B$ 
exponential from the second line in (\ref{thrpnt}) was needed for 
good fits with $t_{max}$ between $17 \sim 21$. Fig.4 plots the 
case $N_\pi=3$ and $N_\pi^\prime=1$.  
The fact that we can fit the data for $t_{max}$ only slightly 
below $t_B$ (the source point for the $B$ meson) with just one or 
two $E_B^{(j)}$ exponentials 
tells us 
that excited $B$ states are highly suppressed in the three-point correlators. 
 We believe this is a physical 
effect indicating that form factors for semileptonic decays of excited 
$B$ mesons (proportional to $A_{jl}$ in (\ref{thrpnt}) with $j > 1$) are 
suppressed relative to ground state form factors. Figs.6 and 7 show 
sample fits to matrix elements of the spatial component of the vector current,
$\langle V^L_k \rangle$.  Again one can go out to $t_{max} \sim 21$ with 
$N_B=1$ for most momenta and with $N_B=2$ for momentum (001).

\vspace{.1in}
\noindent
Our goal is to extract the amplitude $A_{11}$, the groundstate contribution 
to the $C^{(3)}_\mu$.  From $A_{11}$ one can then determine the $B$ meson 
semileptonic decay form factors (see next section).
One consistency check on $A_{11}$ is to verify that its associated 
exponential factors, $e^{-E^{(1)}_\pi t} \, e^{-E^{(1)}_B(t_B-t)}$, 
involve the correct groundstate energies. For $E_B$ this is put 
in by hand through our rescaling in (\ref{rescale}) and the ansatz of 
(\ref{thrpnt}).  Since we are always dealing with zero momentum 
$B$ mesons for which $E^{(1)}_B$ can be determined accurately from 
two-point correlators, we believe this is a sensible way to proceed.
For $E_\pi$, for which we need results for various momenta,
 one possibility is to do simultaneous fits to two- and 
three-point correlators ensuring that the same set of energies appear 
in both correlators \cite{martin}.
  We have opted not to force the $E_\pi^{(l)}$ 
in the two correlators to be equal in this way,  but to do separate 
fits and use consistency between the two independent extractions 
of $E^{(1)}_\pi(p)$ as a check on our fits, on our fit ansatz 
(\ref{thrpnt}), and on our choices for $t_{max}$, $N_B$, $N_\pi^\prime$ etc.
Fig.8 shows groundstate pion energies $E_\pi(p)$ extracted from 
two-point and either $\langle V^L_0 \rangle$ or $\langle V^L_k \rangle$ 
three-point correlators.  One sees good agreement between the different 
determinations.  One also notices that for higher momenta three-point 
correlators provide more accurate energies than two-point correlators.  
Of course, whether this happens or not depends on the smearings used 
in the pion interpolating operator.

\vspace{.1in}
\noindent
In order to include $\Lambda_{QCD}/M$ corrections to the heavy-light 
currents we have looked at the matrix element of 
\be
V^{(1),L}_\mu = \frac{-1}{2M_0}\, \overline{q} \gamma_\mu 
\vec{\gamma} \cdot \vec{\nabla} Q
\ee
(see ref.\cite{pert2} for a complete list of $1/M$ current corrections).  
Figs.9 and 10 show sample fits, similar to Figs.4 and 5, for 
$\langle V^{(1),L}_0 \rangle$. We find that the groundstate contribution 
from $V^{(1),L}_0$ is only a small fraction (1.5 -2.5\%) of the leading 
order contribution from $V^L_0$. 
 Fig.11 shows the ratio between $A_{11}(V^{(1),L}_0)$ and $A_{11}(V^L_0)$. 
We superimpose the $O(\alpha_s/a_sM)$ power law correction that 
must be subtracted from the $\langle V^{(1),L}_0 \rangle$ matrix element 
to obtain the physical $O(\Lambda_{QCD}/M)$ relativistic correction 
\cite{fbscale}.
This is given by the full horizontal line,  the dotted lines 
representing our estimate of uncertainties in the $\alpha_s/a_sM$ power 
law subtractions.  We see that the matrix element is consistent with 
being 100\% power law.  The $1/M$ corrections to the spatial  component
of the vector current, $\langle V^{(1),L}_k 
\rangle$, is found to be even smaller (at the 1\% level) relative to 
the leading order $\langle V^L_k \rangle$.  Since $\langle V_k \rangle$ 
must be proportional to the pion momentum $p_{\pi,k}$ (for $\vec{p}_B=0$), 
whereas  $V^{(1)}_k$ is sensitive mainly to the $b$-quark momentum 
inside the initial $B$ meson at rest,  one would expect 
$\langle V^{(1),L}_k \rangle$ to be very small.  We could not come up with 
a similar plausibility argument as to why the temporal component,
 $\langle V^{(1),L}_0 \rangle$, 
should be small as well. Because the tree-level $1/M$ current matrix 
elements are so small and of the same order of magnitude as 
$O(\alpha_s/a_sM)$ power law corrections,  we have opted not to include 
them in our final analysis.  Uncalculated $O(\alpha_s^2/a_sM)$ 
corrections could 
easily switch the sign of their contributions.  We are dropping terms that 
are 1-3\% of the leading order contributions,  effects that are much 
smaller than the $O(\alpha_s^2)$ systematic errors we will be assigning 
to the present calculation.

\vspace{.1in}
\noindent
 Our results for $\langle V^{(1),L}_\mu \rangle$
 disagree with those in ref.\cite{jlqcd} where a much larger, 10 - 20\%,
contribution from $V^{(1),L}_\mu$ is reported.  The origin of this 
discrepancy is not understood at the present time.  Nevertheless, in 
the next section we will see that our final results for form factors 
agree very well with those of ref.\cite{jlqcd}.  The global fits used 
there to carry out chiral extrapolations must be compensating in part for 
the differences in the $1/M$ current corrections.

\section{Results for Form Factors}

\noindent
The groundstate amplitudes $A_{11}(V^L_\mu)$ extracted from fits to 
three-point correlators in the previous section are related to the 
continuum matrix element of interest as
\be
\label{a11}
\langle \pi(p_\pi) | \, V^\mu \, | B(\vec{p}_B=0)\rangle
 = \frac{A_{11}(V^L_\mu)}
{\sqrt{\xi^{(1)}_\pi \xi^{(1)}_B}} \,2 \, \sqrt{E_\pi M_B} \, Z_{V_\mu}
\ee
$\xi^{(1)}_\pi$ and $\xi^{(1)}_B$ are fixed from $\pi-\pi$ and $B-B$ 
correlators.
\be
\label{picorr}
\sum_{\vec{x}} e^{-i \vec{p} \cdot\vec{x}}
\vev{0|\Phi_\pi(t,\vec{x}) \, \Phi^\dagger_\pi(0) |0} = 
\sum_l \xi^{(l)}_\pi \left [ e^{-E^{(l)}_\pi t} + e^{-E_\pi^{(l)}
(T-t)} \right ]
\ee
\be
\label{bcorr}
\sum_{\vec{x}} 
\vev{0|\Phi_B(t,\vec{x}) \, \Phi^\dagger_B(0) |0} = 
\sum_j \xi^{(j)}_B e^{-E^{(j)}_B t}
\ee

\vspace{.1in}
\noindent
The standard form factors $f_+(q^2)$ and $f_0(q^2)$ are defined through
 ($q^\mu \equiv p_B^\mu - p_\pi^\mu$),
\be
\langle \pi(p_\pi) | \, V^\mu \, | B(p_B)\rangle = 
f_+(q^2) \left [ p_B^\mu + p_\pi^\mu - \frac{M_B^2 - m_\pi^2}{q^2} \, q^\mu 
\right ]
 + f_0(q^2) \frac{M_B^2 - m_\pi^2}{q^2} \, q^\mu
\ee

\vspace{.1in}
\noindent
Following Fermilab \cite{fnal} we have found it convenient to introduce other 
form factors $f_\|$ and $f_\bot$, defined as
\be
\langle \pi(p_\pi) | \, V^\mu \, | B(p_B)\rangle = 
\sqrt{2M_B} [v^\mu f_\| + p_\bot^\mu f_\bot ] .
\ee
with
\begin{eqnarray}
\label{vmu}
v^\mu = \frac{p_B^\mu}{M_B} \quad &\longrightarrow&  \quad (1,\vec{0}) \\
\label{pperp}
p_\bot^\mu = p_\pi^\mu - E_\pi v^\mu \quad  &\longrightarrow& \quad
 (0,\vec{p}_\pi)
\end{eqnarray}
$E_\pi$ is the pion energy in the $B$ meson rest frame and the last 
expressions 
in (\ref{vmu}) and (\ref{pperp}) are similarly the four vectors $v^\mu$ 
and $p^\mu_\bot$ in this frame.
The form factors $f_\|$ and $f_\bot$ are useful since 
(again in the $B$ rest frame) they are simply 
related to the three-point correlators $C^{(3)}_\mu$ for $\mu=0$ and 
$\mu=k$ respectively.  One has
\be
\label{fpara}
f_\| = 
 \frac{A_{11}(V^L_0)}
{\sqrt{\xi^{(1)}_\pi \xi^{(1)}_B}} \, \sqrt{2 E_\pi} \, Z_{V_0}
\ee
and
\be
\label{fperp}
f_\bot = 
 \frac{A_{11}(V^L_k)}
{\sqrt{\xi^{(1)}_\pi \xi^{(1)}_B}} \, \sqrt{2 E_\pi} \, Z_{V_k}
\,/\,p_{\pi,k}
\ee
Once $f_\|$ and $f_\bot$ are determined,  $f_+$ and $f_0$ can then 
be obtained from,
\be
f_+ = \frac{1}{\sqrt{2 M_B}} \, f_\| + \frac{1}{\sqrt{2 M_B}} \,
(M_B - E_\pi) \, f_\bot
\ee
\be
f_0 = \frac{\sqrt{2 M_B}}{(M_B^2 - m_\pi^2)}\, [ (M_B- E_\pi) f_\| 
+ (E_\pi^2 - m_\pi^2) f_\bot ]
\ee
From these formulas one sees that $f_+$ will be dominated by 
$f_\bot$, i.e. by the matrix element of $V_k$, and $f_0$ by $f_\|$ or 
the matrix element of $V_0$.

\vspace{.1in}
\noindent
In Fig.12 we show the form factors $f_+$ and $f_0$ with the
light quark mass fixed at $a_t m = 0.033$, a value which is 
close to the strange quark mass.  The pion momenta span 
 (000),(001),(011),(111),(002) and (112) in units of $2 \pi/L_sa_s$.  One 
sees that statistical errors are reasonable down to about $q^2 = 16 GeV^2$. 
More work is required if one wants to go further away from 
the zero recoil point.

\vspace{.1in}
\noindent
Figs.13 and 14 show chiral extrapolations at fixed pion momentum.
We have tried linear and constant fits to either all 5 data points or 
to just the last 3 points.  The full and dotted lines in Figs.13 and 14 
give some idea of the spread in fit results.  These differences are 
included in the chiral extrapolation systematic errors that we quote. 
With the present statistics it is not sensible to try more sophisticated 
fits.  Much smaller statistical errors and data at smaller light 
quark masses are required to search for chiral log or square root type
behavior. We also believe it is premature to 
try fitting to HQET and/or chiral perturbation theory 
inspired model ansaetze.   In their plot of $f_\|$ and $f_\bot$ versus 
$am_q$ 
the Fermilab collaboration, working 
at smaller quark masses than in this article,
 finds an upward curvature as one decreases the light quark mass 
\cite{fnal}. 
 We cannot 
rule out or verify such behavior with our present data.  
Fig. 15 gives form factors for the physical case $B \rightarrow \pi \;
l \overline{\nu}$.  One sees that errors have increased significantly 
over those in Fig.12.  Furthermore we now  include pion momenta 
only up to (111).  Larger momenta lead to chiral extrapolation errors 
that are too large to make such data points meaningful.

\vspace{.1in}
\noindent
In Fig.16 we compare our results to those by other lattice groups
 \cite{fnal,jlqcd,ape,ukqcd}. 
One sees that agreement for $f_+(q^2)$ is good among all collaborations.  
For $f_0(q^2)$ we agree best with the JLQCD collaboration 
\cite{jlqcd} and are slightly 
below the results of remaining groups.  The Fermilab \cite{fnal} and JLQCD 
collaborations \cite{jlqcd}
are the two other groups that simulate directly at the $b$ quark mass, so 
it is worthwhile making further comparisons with their work. 
  We do so for 
the two form factors $(f_1 + f_2)$ and $f_2$ used by JLQCD, which 
are closely related to $f_\|$ and $f_\bot$.
\be
(f_1 + f_2) = \frac{1}{\sqrt{2}} \, f_\|  \; , \qquad \qquad
f_2 = \frac{E_\pi}{\sqrt{2}} \, f_\bot
\ee
Figs.17 and 18 show comparisons between the three collaborations.  
The form factors are plotted as a function of $E_\pi$, the 
relation between the two variables $q^2$ and $E_\pi$ being 
$q^2 = M_B^2 + m_\pi^2 - 2 M_B E_\pi$.
The Fermilab results for $(f_1+f_2)$ are considerably higher than 
those from the other two collaborations.  The main reason for this 
difference seems to come from the upward curvature, mentioned above, that 
Fermilab sees in their plots such as Figs.13 and 14 of form factors 
versus the light quark (or the pion) mass.  Neither JLQCD nor the 
present work has sufficient accuracy at low enough quark masses to 
see this trend and more calculations are required to resolve 
this issue.  One should note that soft pion theorems, valid in the 
limit $m_\pi \rightarrow 0$ and $\vec{p}_\pi \rightarrow 0$, would 
dictate
\be 
[f_1 + f_2]|_{E_\pi \rightarrow 0} =  \frac{f_B}{2 f_\pi} \sqrt{M_B} .
\ee
The higher Fermilab results in Fig.17 are consistent with this relation 
while JLQCD's and our results are too low.

\section{Summary}
\noindent
We have studied semileptonic $B$ meson decays using highly improved 
gauge and quark actions on anisotropic lattices.  We developed constrained 
fitting methods for analysing three-point correlators and extracting 
groundstate amplitudes in a controlled way.  Our final results 
for form factors agree with previous lattice results.

\vspace{.1in}
\noindent
Our data points in Figs.16, 17 and 18 include  the main systematic 
errors.  Allowing for $8$\% discretization, $4$\% relativistic, 
$8$\%  higher order perturbative and $2$\%  mass tuning 
corrections, we estimate $\sim$12\% systematic errors from all 
sources other than 
quenching and chiral extrapolation.  This is to be compared with 
the $10 - 15$\% chiral extrapolation errors already shown in Fig.15. 
One realizes that accurate semileptonic form factor results will only 
be attainable if uncertainties coming from chiral extrapolations 
are brought under control.
To overcome this obstacle,  we have initiated a program to study 
heavy-light physics with improved staggered light quarks \cite{matt}. 
 Simulations 
can be carried out with much smaller quark masses using this light quark 
action.  
  The experience acquired in the present work and the 
analysis techniques that have been developed for three-point correlators 
will play an important role there.  For instance, with staggered light 
quarks two-point and three-point correlators have time oscillating 
contributions which must be taken into account in fits.
The only way to obtain groundstate contributions to three-point 
correlators will be through fitting them directly, as was done in the 
present article.  Taking ratios of three- and two-point correlators 
will be of no use in simulations with staggered light quarks.
  Other theoretical developments, such as better understanding of 
chiral perturbation theory for staggered fermions \cite{claude} 
and the use of ``Moving 
NRQCD'' \cite{mnrqcd} 
should further aid accurate semileptonic form factor determinations 
in the future.

\vspace{.2in}
\noindent
\acknowledgements

\noindent
This work was supported by the DOE under DE-FG02-91ER40690 and by PPARC 
 and NSF.  Simulations were 
carried out at the Ohio Supercomputer Center and at NERSC.

\noindent
J.S. thanks the Center for Computational Physics, Tsukuba, for 
support and hospitality during the initial stages of this project.
CTH.D. is grateful for a senior fellowship from PPARC.  

\appendix

\section{Constrained Fitting}

\noindent
In this appendix we give some details of our constrained fits to two- and 
three-point correlators.  The general 
approach, within the context of lattice 
simulations, is described in ref.\cite{bayes}.  In lattice simulations 
one typically starts with numerical 
data for some correlator $\overline{G(t)}$, averaged over 
configurations, which 
one wants to fit to a 
theoretical expectation $G_{th}(t)$ to extract energies, amplitudes or 
matrix elements.  Examples of $G_{th}(t)$ would be the 
RHS's of (\ref{thrpnt}), (\ref{picorr}) or (\ref{bcorr}), which we 
can generically write as,
\be
\label{gth}
G_{th}(t) = \sum_n A_n e^{-E_n t} .
\ee
Denoting the fit parameters $A_n$ and $E_n$ collectively as $\alpha_j$, one 
has 
\be
G_{th}(t) = G_{th}(t,\{\alpha_j\}) .
\ee
Conventional fits are carried out by minimizing the $\chi^2$,
\be
\chi^2(\{\alpha_j\}) = \sum_{t,t^\prime} 
[\overline{G(t)} - G_{th}(t,\{\alpha_j\})] \; \sigma^{-1}_{t,t^\prime} \;
[\overline{G(t^\prime)} - G_{th}(t^\prime,\{\alpha_j\})] 
\ee
with respect to the parameters $\{\alpha_j\}$.  
$\sigma^{-1}$ is the inverse of the correlation matrix,
\be
\sigma_{t,t^\prime} = \overline{G(t) G(t^\prime)} - \overline{G(t)} \;
\overline{G(t^\prime)}
\ee
Depending on the quality of the data, only a few low lying energies and 
amplitudes will be constrained by the data.  If one includes too many 
terms in (\ref{gth}), the unconstrained $E_n$'s and $A_n$'s for higher 
$n$ can wander all over the place and start to destabilize the fits.

\vspace{.1in}
\noindent
``Constrained fits'' were proposed in ref.\cite{bayes}
 to get around this problem.
One augments the conventional $\chi^2$ with a term, $\chi^2_{prior}$, which 
prevents fit parameters that are not constrained by the data from 
taking on ``unreasonable'' unphysical values.
\be
\label{chiaug}
\chi^2 \longrightarrow \chi^2_{aug} \equiv \chi^2 + \chi^2_{prior},
\ee
with
\be
\chi^2_{prior} \equiv \sum_j  \frac{(\alpha_j - \tilde{\alpha}_j)^2}
{\tilde{\sigma}^2_j} .
\ee
In this scheme each parameter $\alpha_j$ has its set of ``priors'', 
$\tilde{\alpha}_j$ and $\tilde{\sigma}_j$,  and $\chi^2_{aug}$ 
is designed to favor $\alpha_j$ values in the range $\tilde{\alpha}_j 
\pm \tilde{\sigma}_j$. The replacement of eq.(\ref{chiaug}) can be 
justified within the framework of Bayesian statistics and Bayes' theorem. 
It implies using 
a Gaussian apriori distribution for the parameters $\{\alpha_j\}$.
 For parameters $\alpha_j$ that 
are determined by the data, adding $\chi^2_{prior}$ has minimal effect 
on the final fit value, 
as long as $\tilde{\sigma}_j$ is not made too small.  In the present work 
we always set $\tilde{\sigma}_j = \tilde{\alpha}_j$. One is 
dealing with very wide Gaussians and hence very unrestrictive priors.
We have checked that changing $\tilde{\sigma}$ to 75\% or 50\% or even 
25\% of 
$\tilde{\alpha}_j$  does not change results for data-determined 
fit parameters.  Choices for the central values $\tilde{\alpha}_j$ are 
made based on preliminary fits and physics input about typical level 
splittings in the system under study.  Again, if the $\tilde{\sigma}_j$'s 
are wide enough final results for data-determined parameters are 
not sensitive to precise values of the $\tilde{\alpha}_j$'s.

\vspace{.1in}
\noindent
The method is best illustrated by an explicit example.  Table II lists 
priors used in fits to pion two-point correlators for our 
next to lightest quark mass.  Figs.19 and 20 show fit results for the 
groundstate energy as a function of the number of cosh's for several 
pion momenta.  The numbers below the data points show the 
$\chi^2/d.o.f.$ for the fits.  The fit range is shown on the top left 
corner of the plots.  One sees that good fits are obtained for 
$N_{cosh} \ge 4$.  The fancy stars show bootstrap fit results (bootstrap 
methods within constrained fits are discussed below).  
The priors $\tilde{E}_1$ and $\tilde{A}_1$ and $\tilde{E}_2$ 
were chosen from preliminary 
fits or by looking at effective mass plots.  We have 
checked that changing them by factors of 2 moves fit results 
for groundstate energies and amplitudes by much 
less than their fit errors. 
(precise definition of fit errors will be given 
below when we discuss bootstrap methods and bootstrap errors).
In Fig.21 we show what happens if one changes the priors for the 
higher states ($n >2$) from $[\tilde{E}_n = \tilde{E}_1 + (n-1)
 \times 0.3] $ to 
 $ [\tilde{E}_1 + (n-1) \times 0.2] $, or from $\tilde{A}_n = 0.05$ to 
$\tilde{A}_n = 0.08$.

\vspace{.1in}
\noindent
One sees from Figs.19 and 20 that once sufficient number of exponentials 
(cosh's) are included, fit results stabilize.  We then fix $N_{cosh}$ and 
carry out bootstrap fits for our final analysis.  For instance, for 
pion two-point correlators we choose $N_{cosh} = 4$.

\vspace{.1in}
\noindent
In a bootstrap analysis involving constrained fits one first creates 
a certain number  (we choose nboot=200) of bootstrap ensembles in the 
usual way. For each bootstrap ensemble a different prior value
 $\tilde{\alpha}_j$ is picked at random for each $j$ according to 
a Gaussian distribution about a central value $\tilde{\alpha}_
{j,0}$ with width $\tilde{\sigma}_j$.  In bootstrap fits, Table II should 
be viewed as giving values for $\{\tilde{\alpha}_{j,0}\}$ rather 
than for  $\{\tilde{\alpha}_j\}$ and we set $\tilde{\sigma}_j = 
\tilde{\alpha}_{j,0}$.  Fits are carried out for each of the nboot bootstrap
ensembles using $\chi^2_{aug}$ with the $\{\tilde{\alpha}_j\}$ for 
that ensemble.  In order to get a bootstrap average and bootstrap errors 
one sorts the nboot fit values according to size and discards the 
top and bottom 16\%.  We take the average of the remaining 68\% as 
our bootstrap average and one half of the difference between the 
largest and smallest values within the 68\% as our bootstrap error.  
The fancy squares in Figs.19 and 20 give bootstrap results calculated 
this way.  One sees very good agreement between bootstrap and non-bootstrap 
single fits.  For the latter, errors are calculated from the square root
of the diagonal elements 
 of the covariance matrix $C$, defined through
\be
(C^{-1})_{ij} \equiv \frac{1}{2}\, \frac{\partial^2 \chi^2_{aug}}
{\partial \alpha_i \, \partial \alpha_j}.
\ee
Given the very different definition of errors and the fact that in the
bootstrap fits very different priors are being used compared to 
in the single fits (where $\{\tilde{\alpha}_j\} \equiv
 \{\tilde{\alpha}_{j,0}\}$),  we find the consistency between the two 
types of fits very reassuring. 
In Figs.22 and 23 we show fit results for the groundstate amplitudes, 
$A_{11}$, contributing to the $\langle V_0 \rangle$ threepoint 
correlator (see eqns.(\ref{thrpnt}) and (\ref{a11}) for definition of 
$A_{11}$).
  The $N_\pi=3$ fits 
are those that went into the plots of Figs.4 and 5.  One again sees 
good agreement between single and bootstrap fits.

\section{One-loop Perturbative Matching}

\noindent
In this appendix we summarize the perturbative calculations necessary 
to match the NRQCD/D234 heavy-light vector current to its continuum QCD 
counterpart at the $O(\alpha_s)$ and $O(\alpha_s/a_sM)$ level.  The formalism 
is described in detail in ref.\cite{pert2}. 
 We have generalized those calculations 
to include anisotropic lattices, improved glue and a more highly improved 
light quark action.  The D234 one-loop 
self energy corrections have already been calculated in ref.\cite{pert}
 for these 
more complicated lattices and glue actions. 
In the present work we do not include $O(a_s \alpha_s)$ or 
$O(\alpha_s \frac{\Lambda_{QCD}}{M})$ terms in the action or in the 
currents.  Hence, in the notation of ref.\cite{pert2,fbscale}, 
only the $\zeta_{00}$ 
and $\zeta_{10}$ elements of the mixing matrix are required, in addition 
to the heavy quark self energy.  The relation between current matrix 
elements $\langle V_\mu \rangle$ in continuum QCD and the matrix elements 
$\langle V^L_\mu \rangle$ evaluated on the lattice, is
given to this order by,
\be
\langle V_\mu \rangle = \frac{1}{\sqrt{Z_q^{(0)}}} \; \left \{
 [ 1 + \alpha_s \tilde{\rho}_{0,\mu} ] \; \langle V^L_\mu \rangle  
+ \langle V^{(1),L}_\mu \rangle_{sub} \; \right \},
\ee
with
\be
\tilde{\rho}_{0,\mu} = B_\mu -\frac{1}{2} (C_q + C_Q) - \zeta_{00,\mu}
\ee
and
\be
\langle V^{(1),L}_\mu \rangle_{sub}  =  
 \langle V^{(1),L}_\mu \rangle - \alpha_s \zeta_{10,\mu} \, \langle V^L_\mu 
\rangle.
\ee
The second term proportional to $\zeta_{10,\mu}$
is the $O(\alpha_s/a_sM)$  power law subtraction term plotted in Fig.11.
$C_q$ and $C_Q$ are the one-loop light and heavy quark wave function 
renormalizations, $Z_q^{(0)}$ is the tree-level light quark 
wave function renormalization,
\be
Z_q^{(0)} = \frac{1}{\sqrt{(a_t m)^2 + 2 (a_t m) \chi + 1}}, 
\ee 
and $B_\mu$ is given by,
\be
B_0 = \frac{1}{\pi} \left [ - \frac{1}{4} + ln(a_sM) \right ],
\qquad \qquad 
B_k = \frac{1}{\pi} \left [ - \frac{11}{12} + ln(a_sM) \right ].
\ee
In Table III we list one-loop results for $C_q$, $C_Q$, $\zeta_{00,\mu}$ 
and $\zeta_{10,\mu}$ for $a_sM = 4.0$ and massless light quarks.  We work 
 in general gauge ($\alpha_g=1 $ and $\alpha_g=0$ 
correspond to Feynman and Landau gauges respectively) and use
gauge invariance of $\zeta_{10,\mu}$ and the combination 
$[\frac{1}{2}(C_q+C_Q) + \zeta_{00,\mu}]$ as checks on our calculations. 
In Table III we  present only the IR finite parts of $C_q$, $C_Q$ and 
$\zeta_{00,\mu}$.  The IR divergent pieces cancel between the lattice 
and continuum parts of the matching calculation.  The Landau gauge 
results have smaller numerical integration errors since both the 
light quark wave function renormalization and the heavy-light 
vertex correction are IR finite in this gauge.  

\vspace{.1in}
\noindent
For reasons described in the text, we do not include $\langle V^{(1)}_\mu 
\rangle_{sub}$ in our final results.  The matching factors $Z_{V_\mu}$ 
of eqs.(\ref{a11}), (\ref{fpara}) and (\ref{fperp}) are then given by,
\be
Z_{V_\mu} = \frac{1}{\sqrt{Z^{(0)}_q}} \, [ 1 + \alpha_s \tilde{\rho}_{0,\mu}]
\ee
We have used $\alpha_s \approx 0.25(5)$ in our perturbative matching.  These 
values are close to $\alpha_V(2/a_s)$ estimated on isotropic lattices 
with unimproved glue. Systematic errors assigned to higher order 
perturbative corrections should cover this uncertainty in $\alpha_s$.

%
%

\begin{table}
\caption{Simulation Details.  
  }
\begin{center}
\begin{tabular}{cccc}
& lattice size      &  $ 12^3 \times 48 $ & \\
&\# configs         &   199              &\\
&$\beta$            &   2.4               &\\
&Landau link $u_0$   
         &  $u_s$=0.7868  $\;u_t$=0.9771   &\\
&$\chi_0$         &  3.0            & \\
&$\chi = a_s /a_t$   &  2.71(3)     & \\
&$C_0$               &    0.94      &   \\
&$a_s^{-1}$       &   1.20(5) GeV   & \\
&$a_t m$           &   0.023 - 0.043 &          \\
&$P/V$             &   0.62 - 0.76    & \\
&$a_sM_0$       &  4.0   & \\
\end{tabular}
\end{center}
\end{table}

\begin{table}
\caption{Priors used in pion correlators for $a_t m =0.028$
  }
\begin{center}
\begin{tabular}{cr|ccc}
& &  $\tilde{E}_n  = \tilde{\sigma}_{E_n}$
 &  $\tilde{A}_n  = \tilde{\sigma}_{A_n}$  &\\
\hline
&\underline{$n=1$} $\qquad \quad$ && &  \\
&mom = (000) &  0.20   & 0.07 & \\
&      (001) &  0.28   & 0.04 &  \\
&      (011) &  0.34   & 0.03 &  \\
&      (111) &  0.40   & 0.02 &  \\
&      (002) &  0.50   & 0.02 &  \\
&      (112) &  0.54   & 0.02 &  \\
\hline
&   &&&  \\
& \underline{$n > 1$}$ \qquad \quad$ 
       &   $\tilde{E}_1 + (n-1) \times 0.3$  &  0.05  & \\
&&&&  \\
\end{tabular}
\end{center}
\end{table}

\begin{table}
\caption{One-loop perturbative coefficients for $a_s M = 4.0$, 
massless light quarks and $\chi = 2.71$.  The coefficients $C_Q$, $C_q$, 
$\zeta_{00,\mu}$, $\zeta_{10,\mu}$ and $\tilde{\rho}_{0,\mu}$ are defined in 
Appendix B.  $\alpha_g$ is the gauge parameter.
  }
\begin{center}
\begin{tabular}{c|cc|cc}
&& $V_0 \qquad \qquad \qquad$ & & $V_k \qquad \qquad \qquad$ \\
\hline
& $\alpha_g = 1$     &  $\alpha_g = 0$
& $\alpha_g = 1$     &  $\alpha_g = 0$   \\
\hline
&&&&  \\
$C_Q$      & 0.020(3)   &  0.520(3)   &&  \\
$C_q$      & -0.066(3)   &  0.435(1)  &&  \\
&&&&  \\
$\zeta_{00,\mu}$  &  0.629(1)   & 0.1285   & 0.506(1)  &  0.0070  \\
&&&&  \\
$\zeta_{10,\mu}$  &  -0.096  &  -0.096  &  0.054  &  0.054  \\
&&&&  \\
\hline
&&&&  \\
$[\frac{1}{2}(C_Q + C_q)$   &&&&  \\
$ \qquad + \, \zeta_{00,\mu}]$  &  0.606(3)  &  0.606(2)  &  0.483(3) 
 & 0.485(2) \\
&&&&  \\
$\tilde{\rho}_{0,\mu}$  &  -0.244(3)  &  -0.244(2)  & -0.334(3)  &  -0.332(2)\\
&&&&  \\
\end{tabular}
\end{center}
\end{table}


\begin{figure}
\begin{center}
\epsfysize=7.in
\centerline{\epsfbox{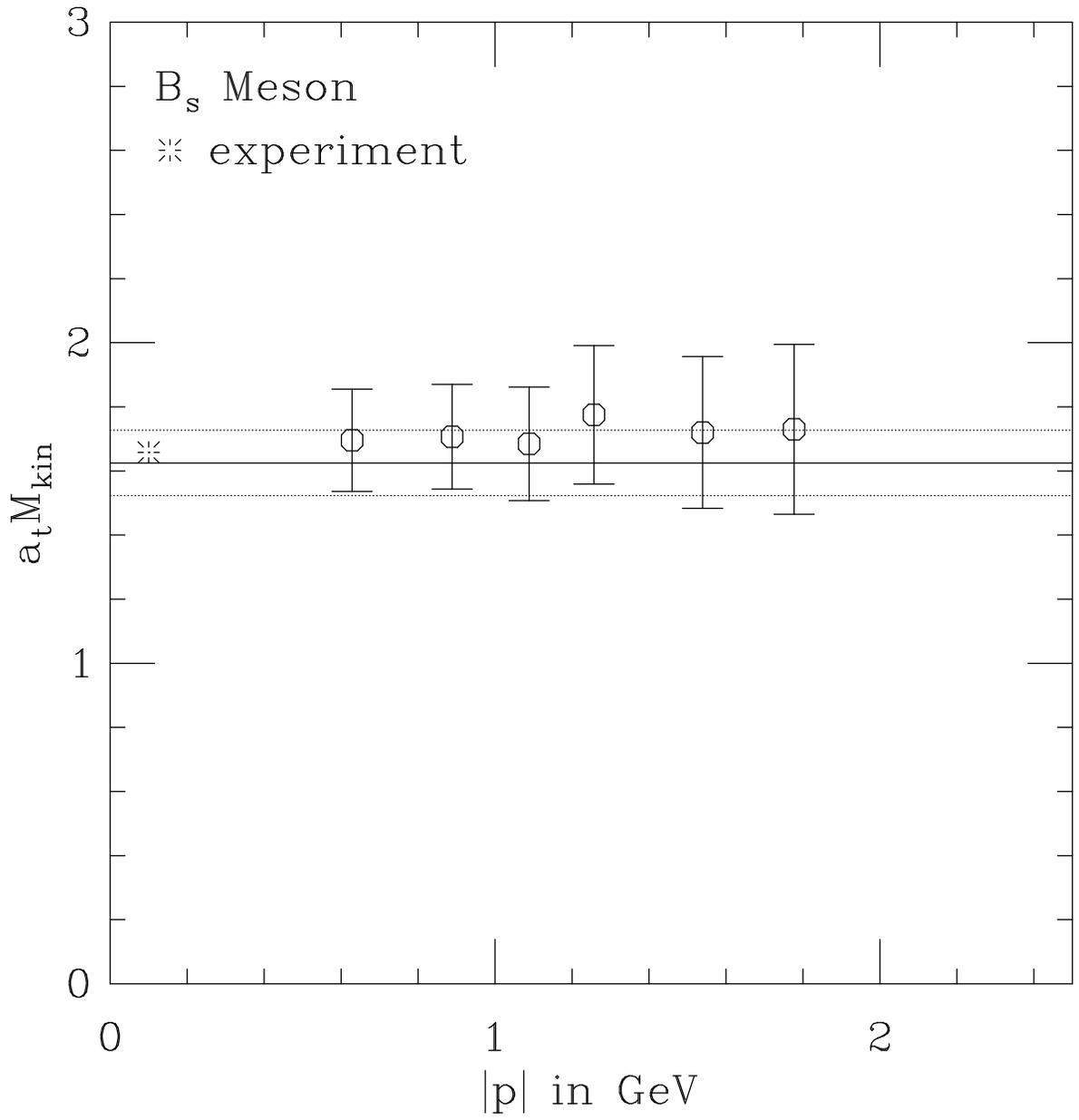}}
\end{center}
\caption{$a_t M_{kin}$ derived from correlators with different momenta.
The full horizontal line gives the one-loop perturbative estimate. 
The two dotted horizontal lines indicate perturbative errors. }

\end{figure}

\begin{figure}
\begin{center}
\epsfysize=7.in
\centerline{\epsfbox{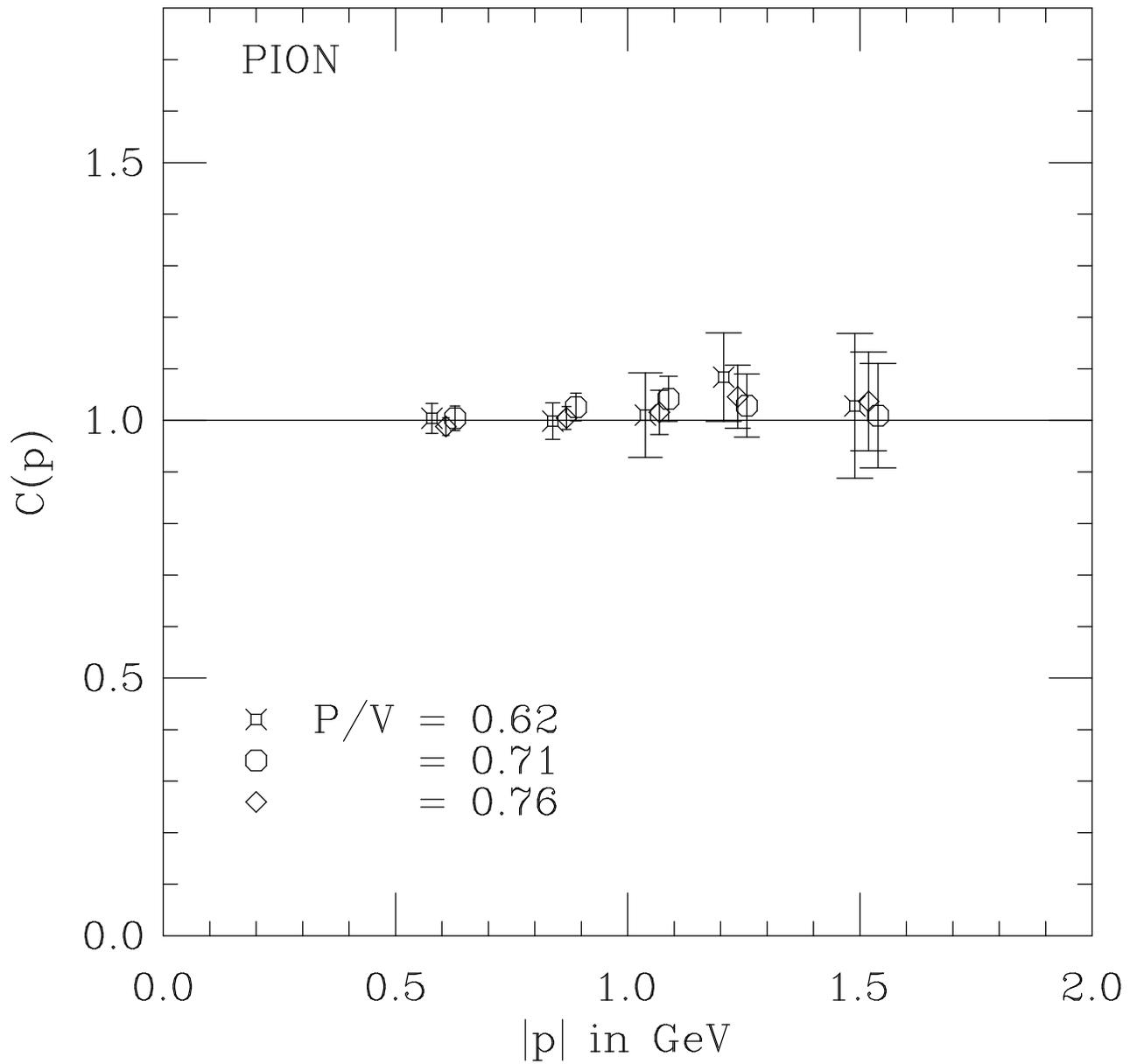}}
\end{center}
\caption{$C(p)$ versus the pion momentum for three light quark masses. }
\end{figure}

\begin{figure}
\begin{center}
\epsfysize=7.in
\centerline{\epsfbox{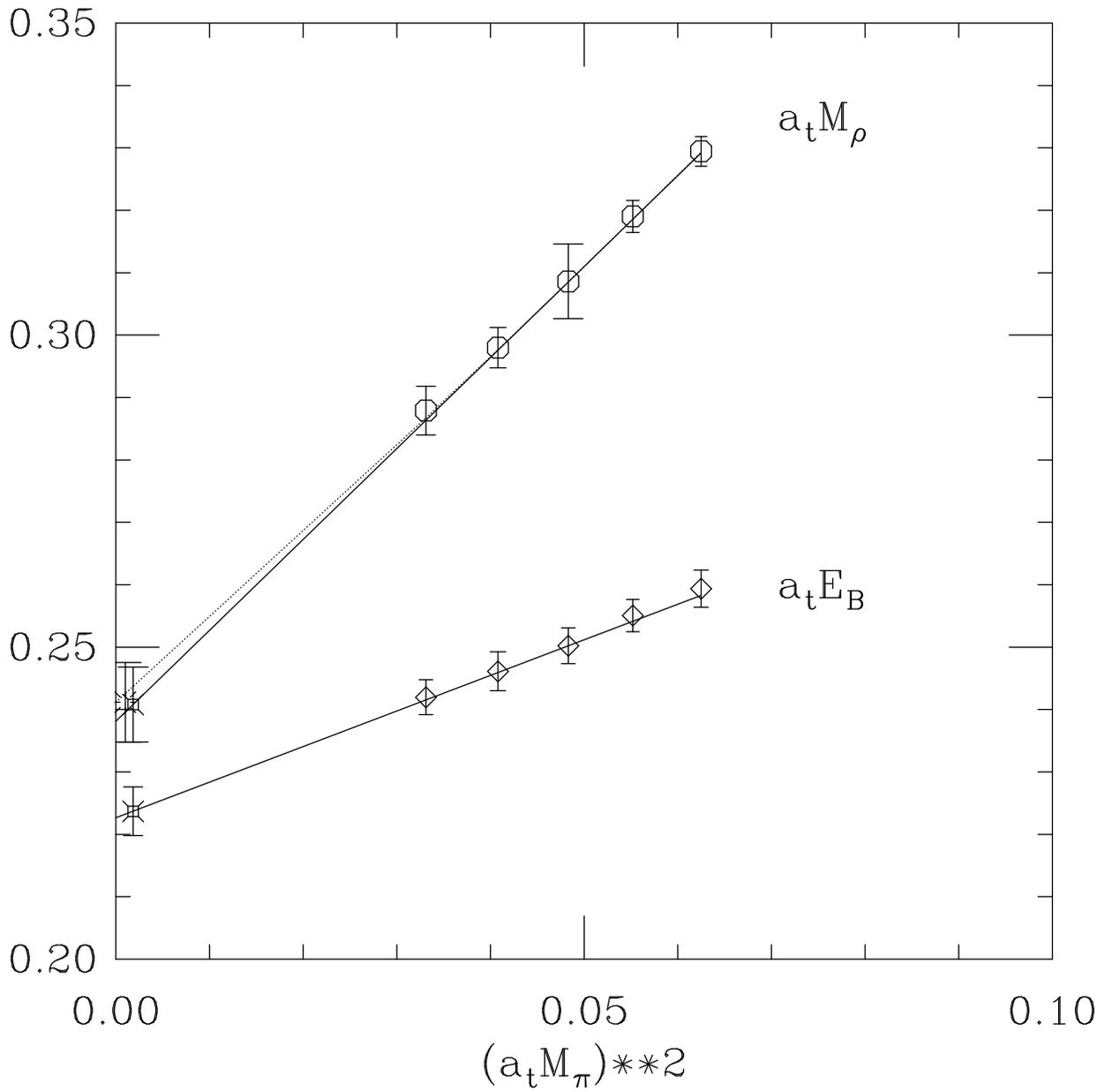}}
\end{center}
\caption{Chiral extrapolation of the $\rho$ mass and of $E_B(0)$. 
For $a_t m_\rho$ both linear (full line) and quadratic (dotted line) 
extrapolations are shown.
}
\end{figure}

\begin{figure}
\centerline{
\ewxy{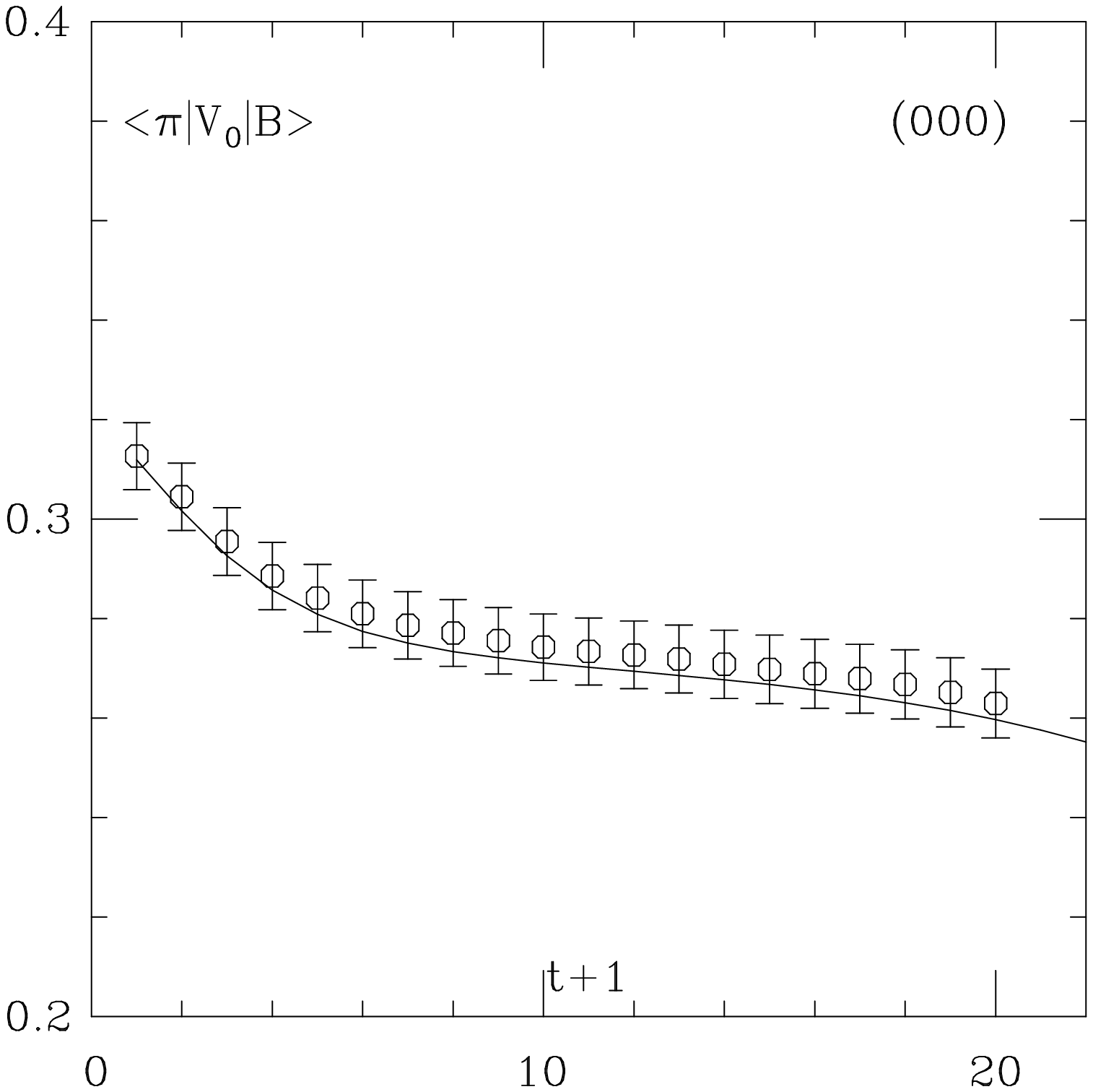}{90mm}
\ewxy{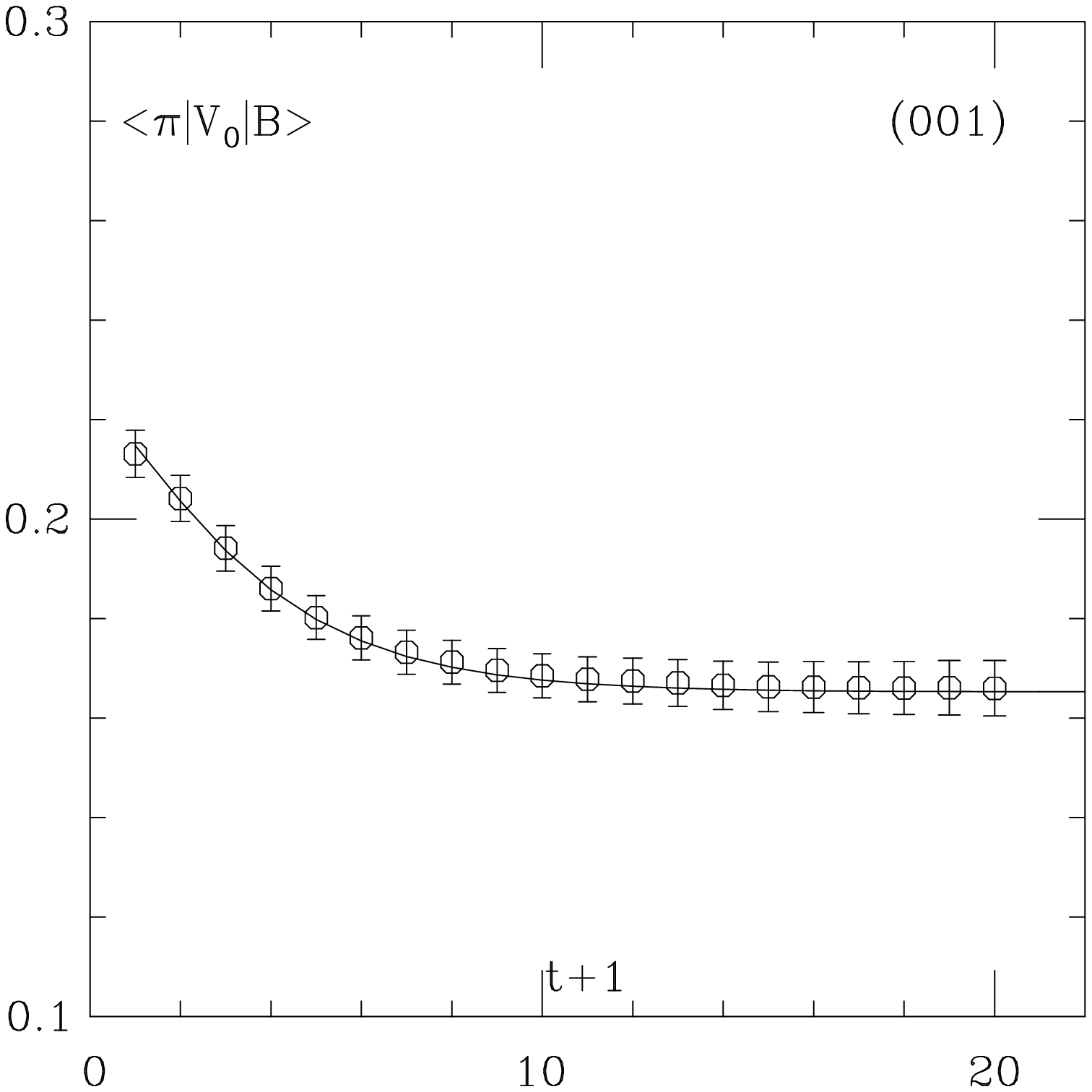}{90mm}
}
\caption{Fits to 3-point correlators.
All fits have 3 exponentials 
coming in from the left and 1 or 2 exponentials from the right.  Both 
the fit and the data have been multiplied by $e^{E^{(1)}_\pi\,t}
 e^{E^{(1)}_B
(t_B - t)}$ for presentation purposes. Results are shown for strange type 
light quarks, i.e. for the third (middle) light quark mass out of a 
total of 5.  In the upper right corner we show the momentum of the pion.}
\end{figure}

\begin{figure}
\centerline{
\ewxy{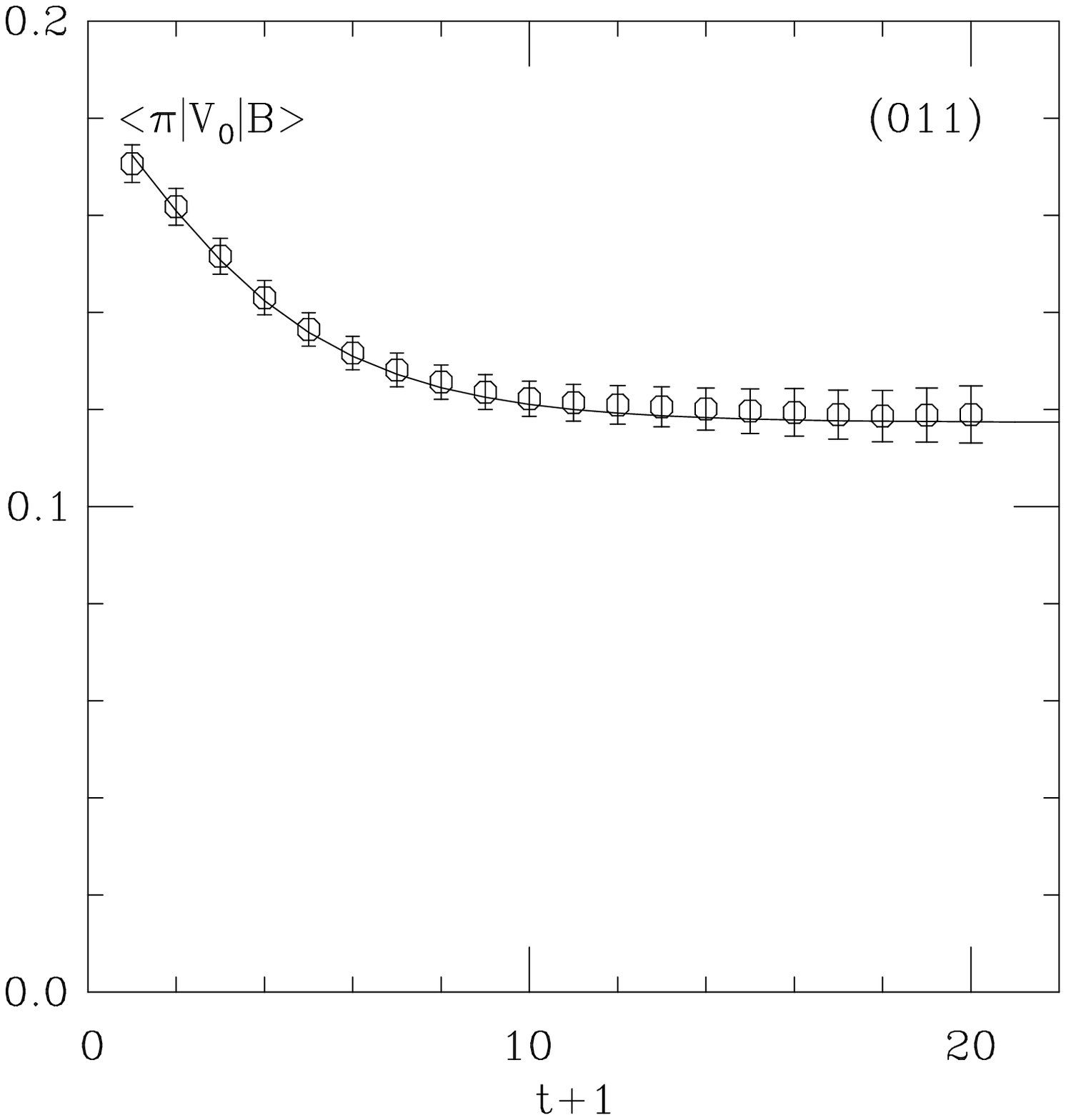}{90mm}
\ewxy{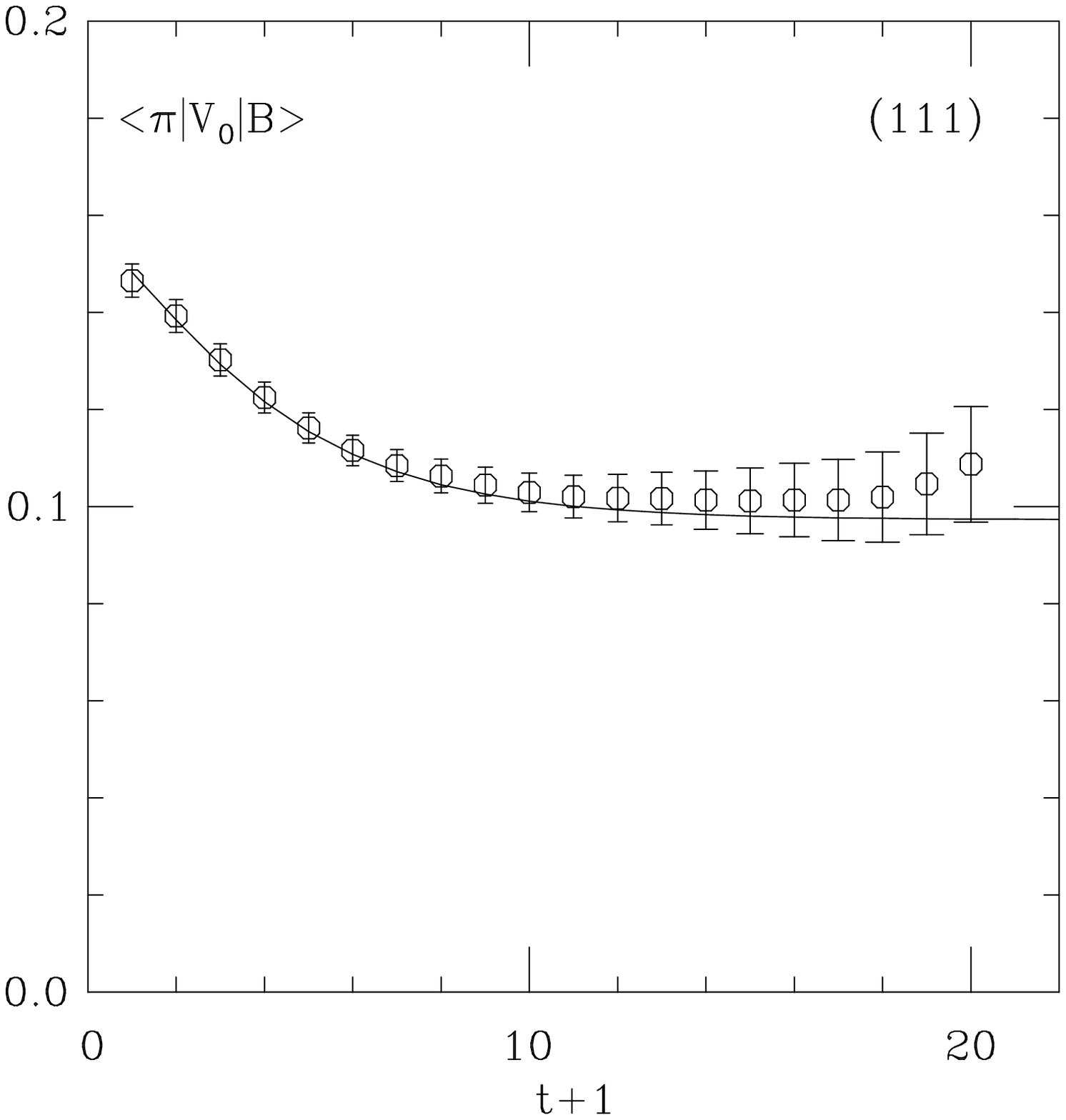}{90mm}
}
\caption{same as Fig.4 for momenta (011) and (111).}
\end{figure}

\newpage
\begin{figure}
\centerline{
\ewxy{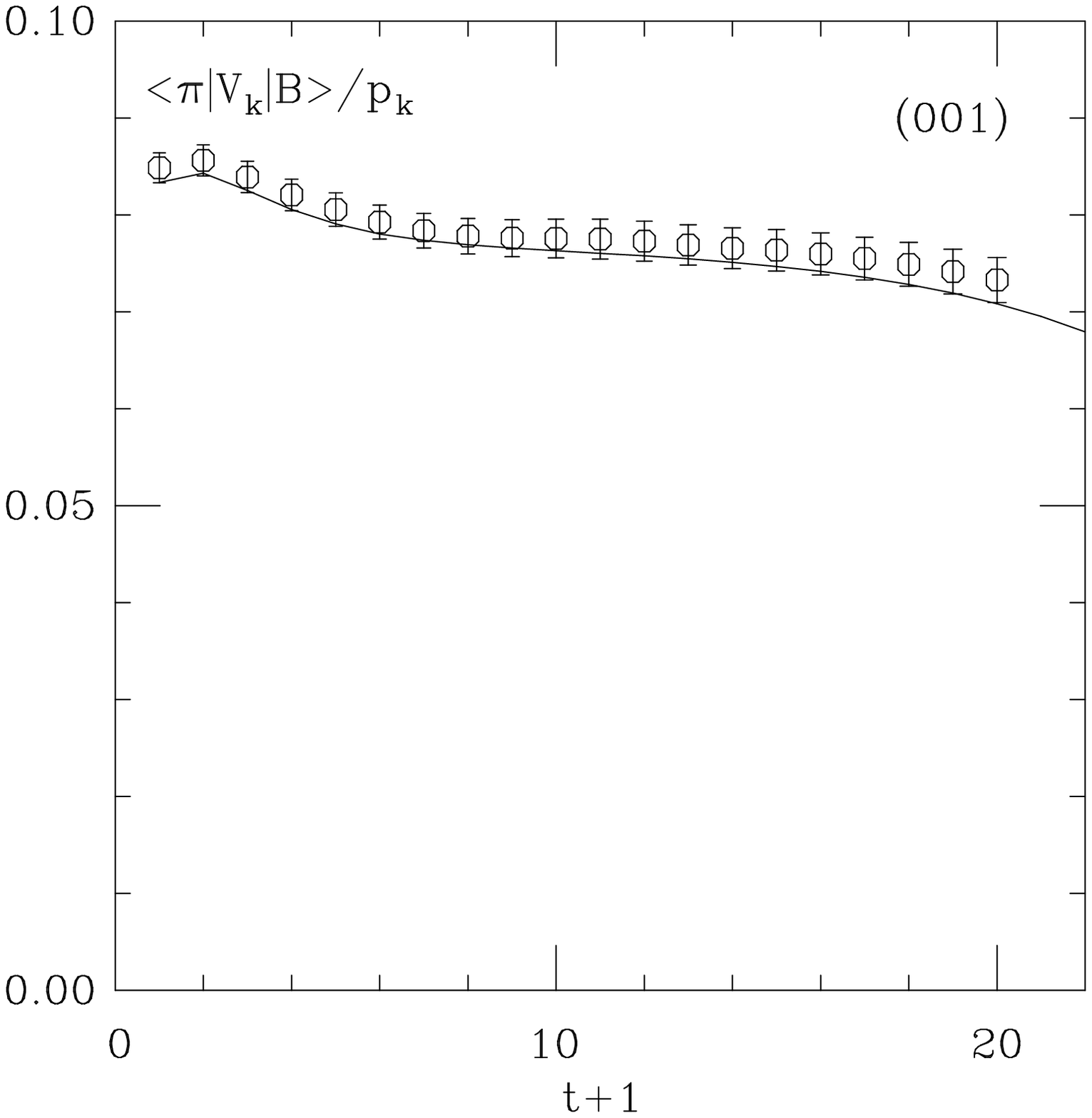}{90mm}
\ewxy{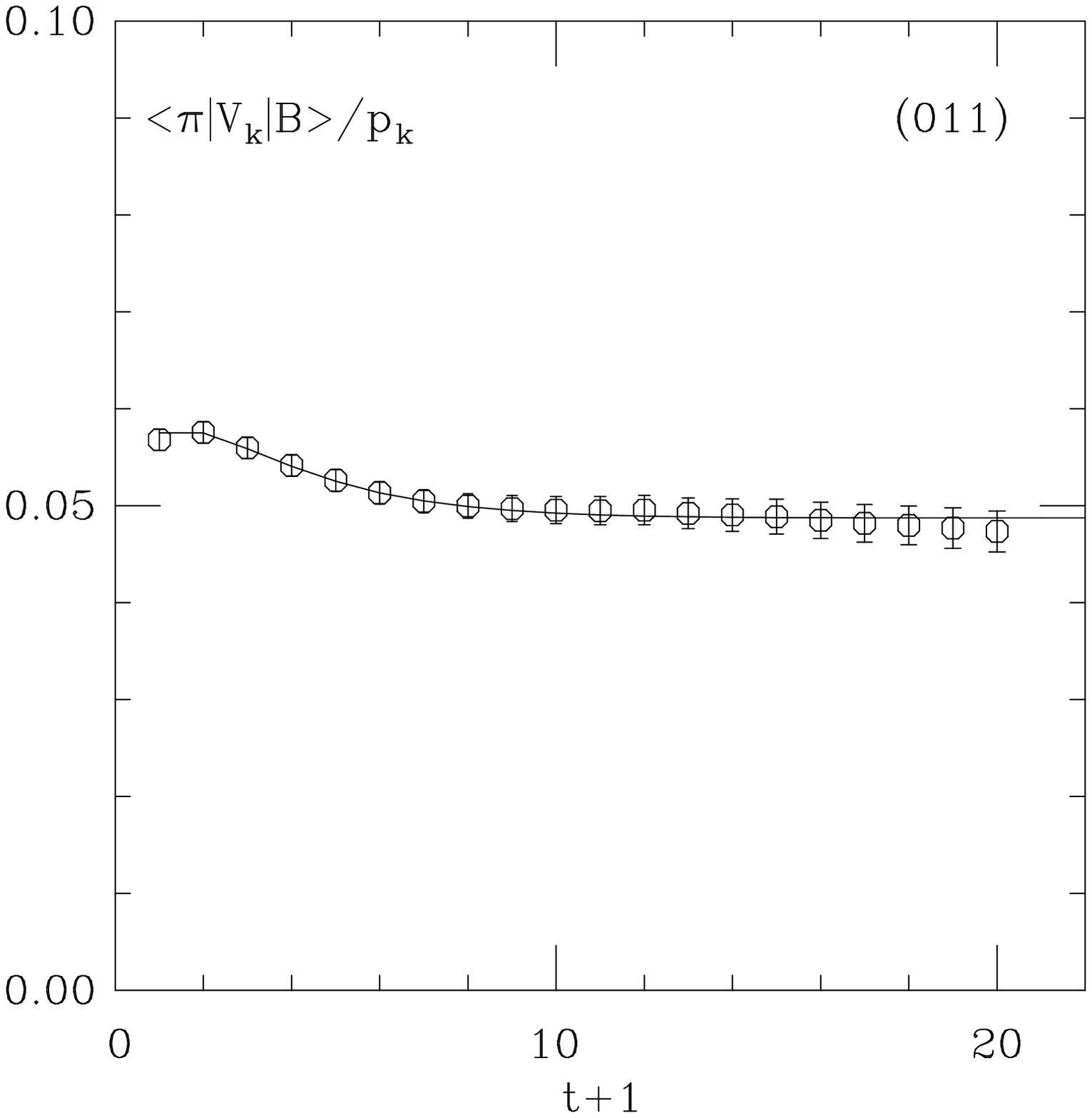}{90mm}
}
\caption{same as Fig.4 for $V_k$ }
\end{figure}

\begin{figure}
\centerline{
\ewxy{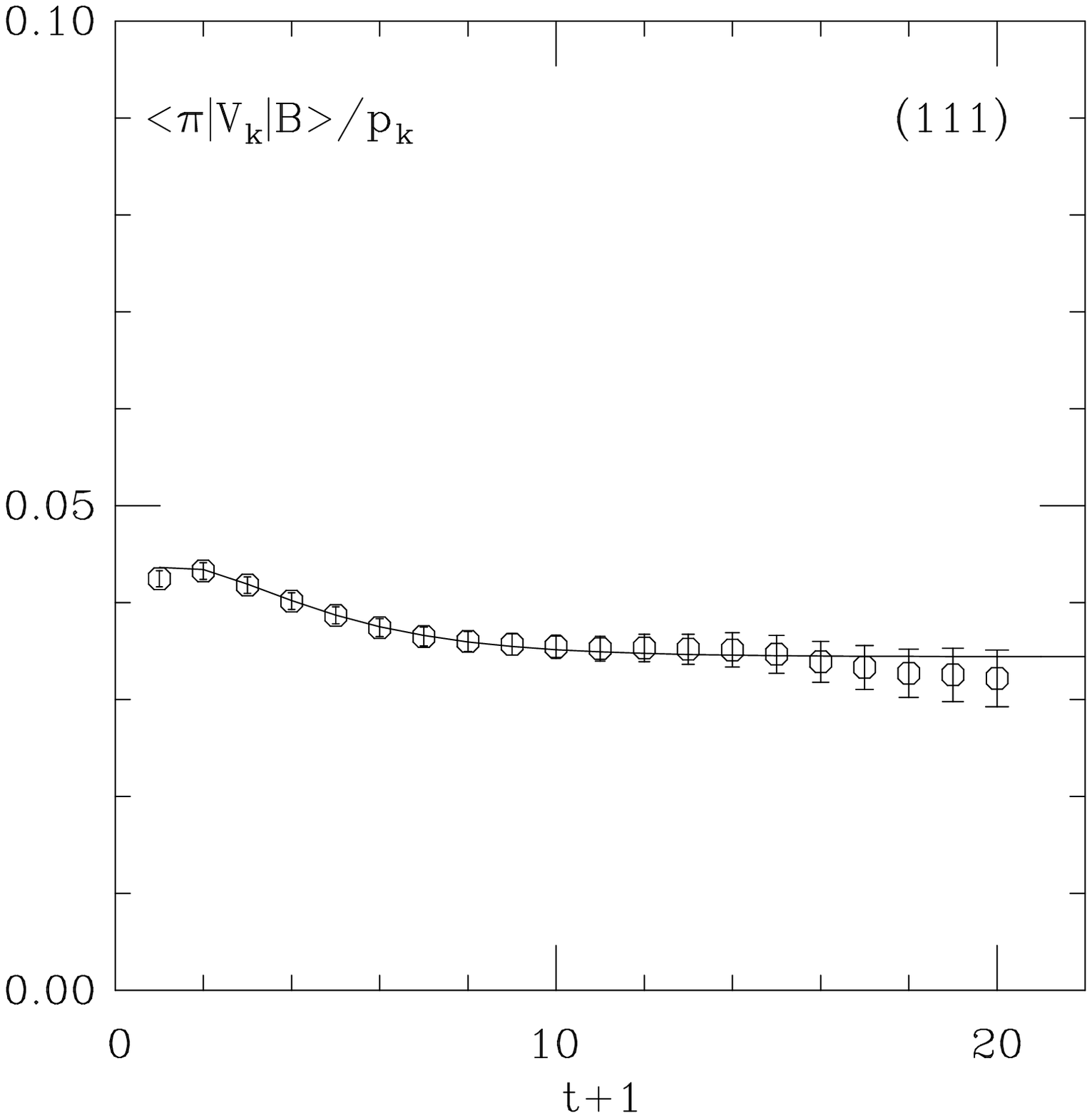}{90mm}
\ewxy{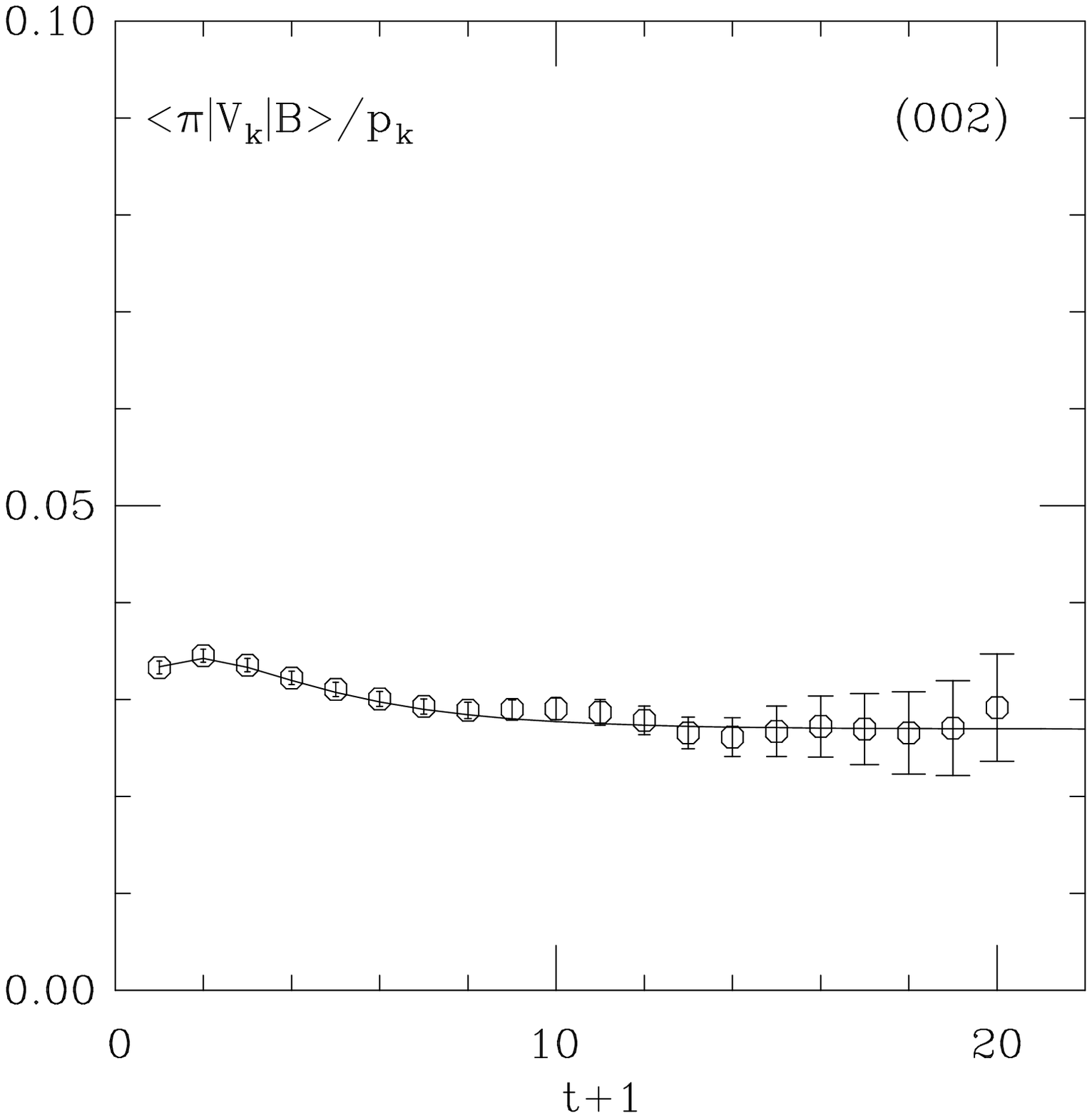}{90mm}
}
\caption{same as Fig.4 for $V_k$ }
\end{figure}

\begin{figure}
\begin{center}
\epsfysize=7.in
\centerline{\epsfbox{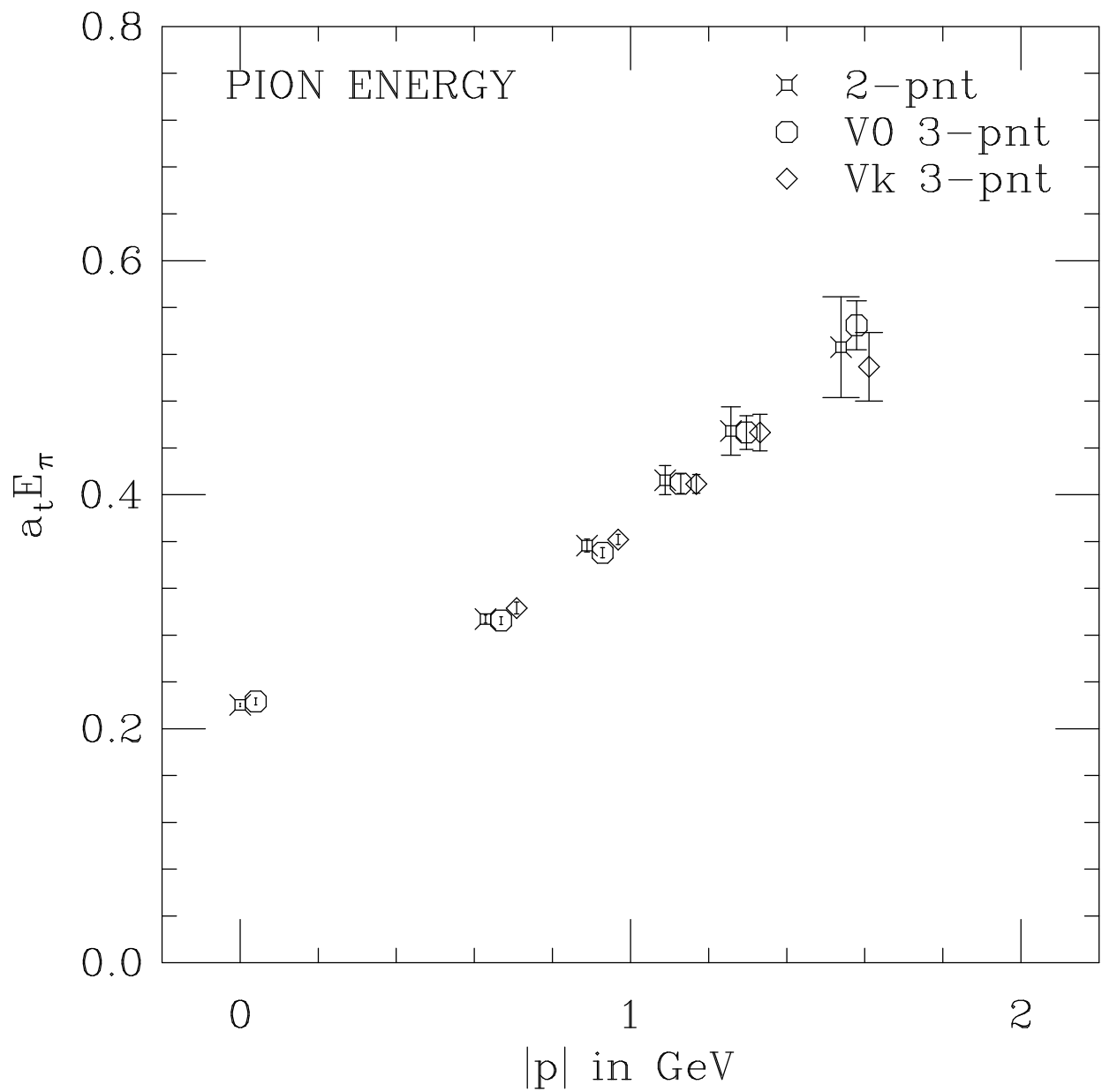}}
\end{center}
\caption{Consistency test for pion energies extracted from different 
correlators }
\end{figure}

\newpage
\begin{figure}
\centerline{
\ewxy{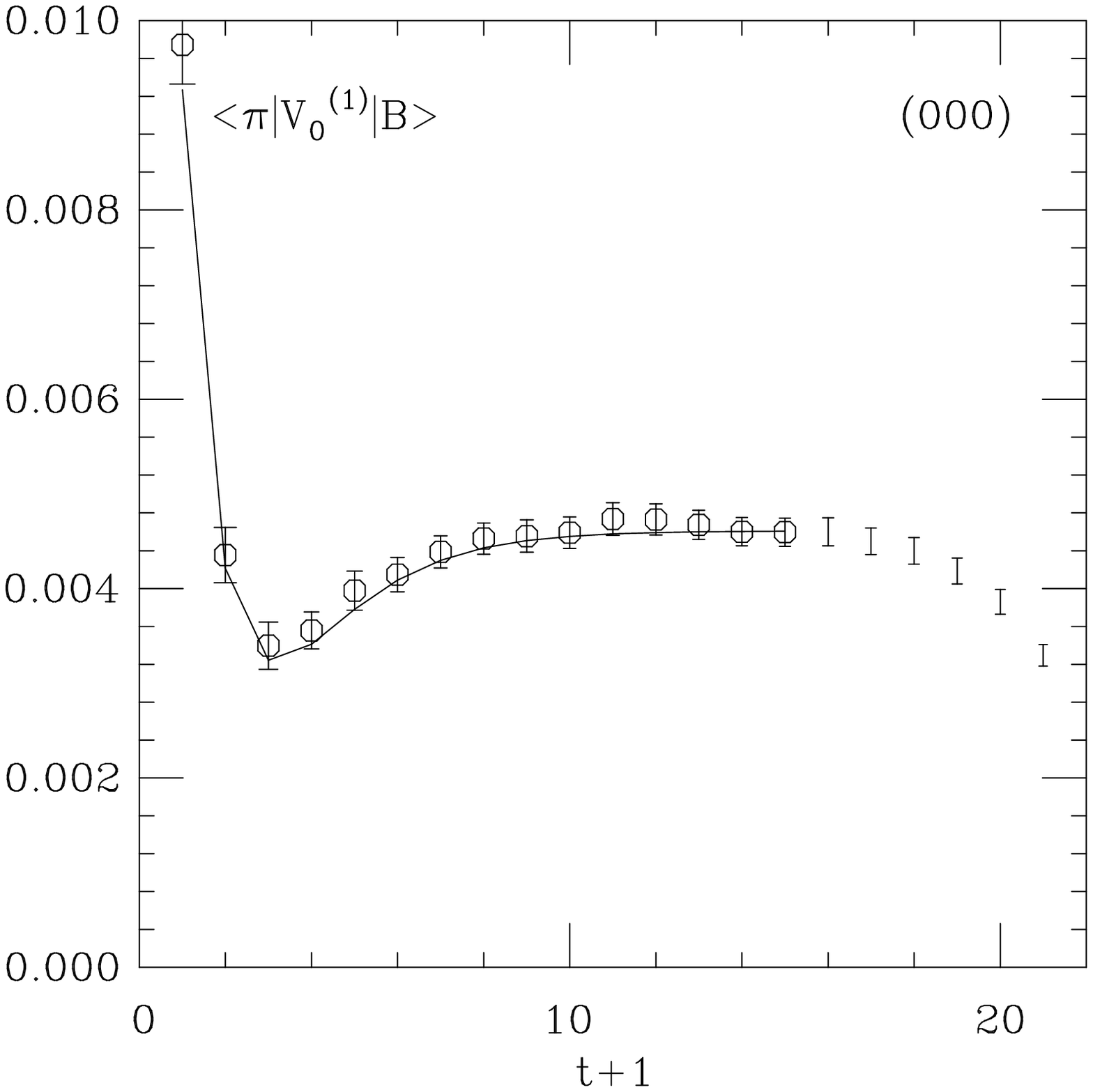}{90mm}
\ewxy{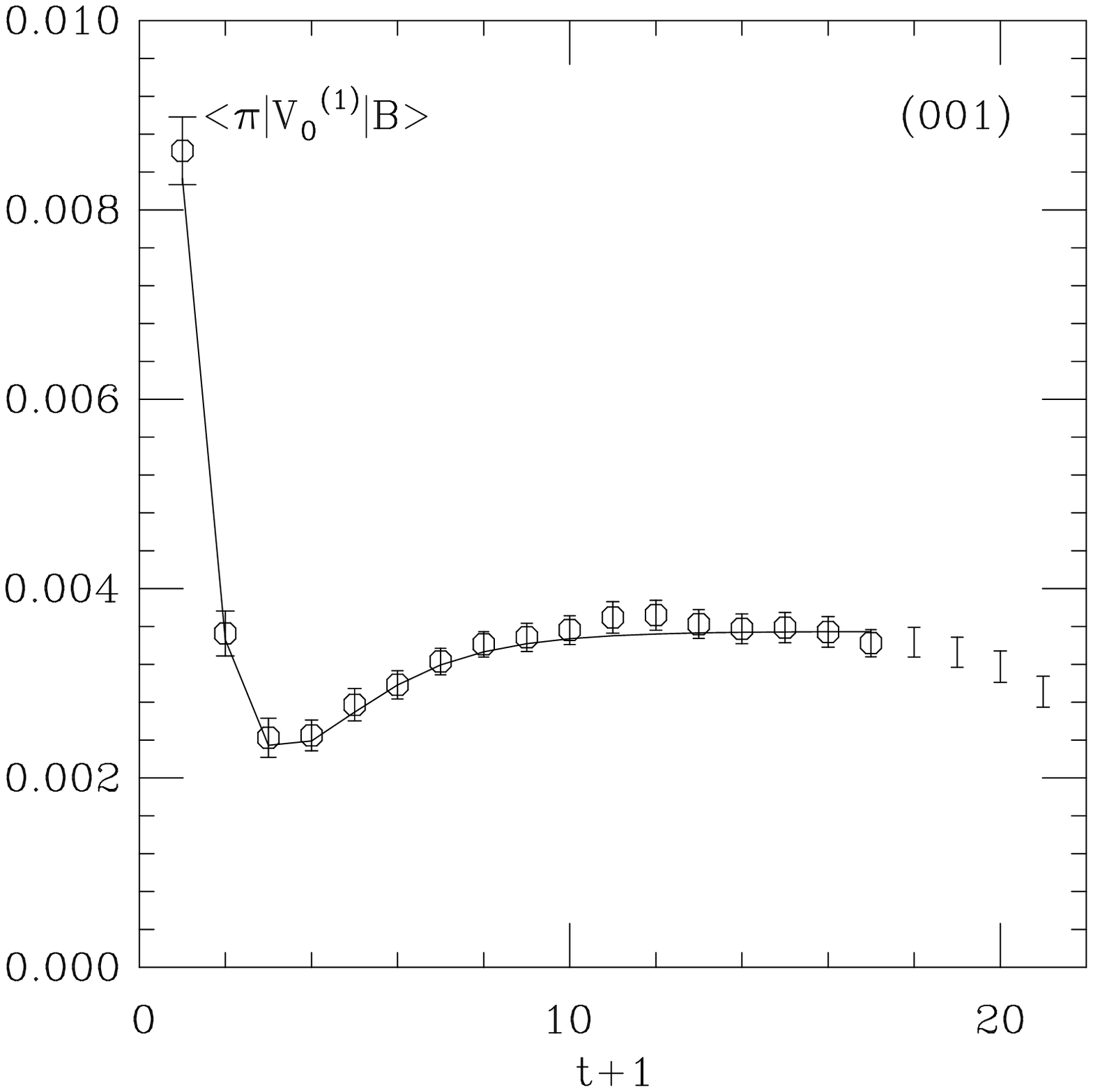}{90mm}
}
\caption{same as Fig.4 for $V_0^{(1)}$, the tree-level 1/M current
 correction }
\end{figure}

\begin{figure}
\centerline{
\ewxy{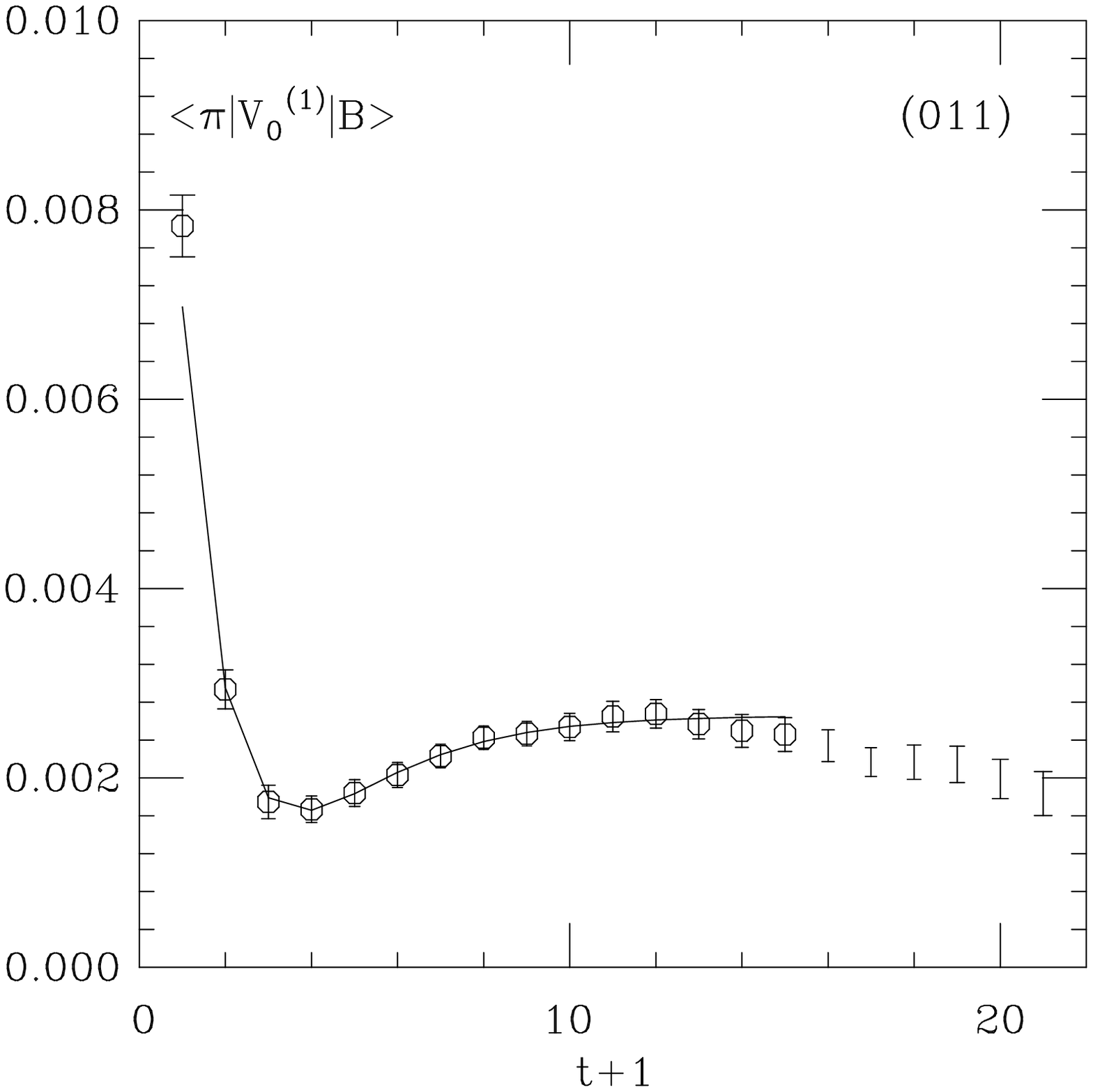}{90mm}
\ewxy{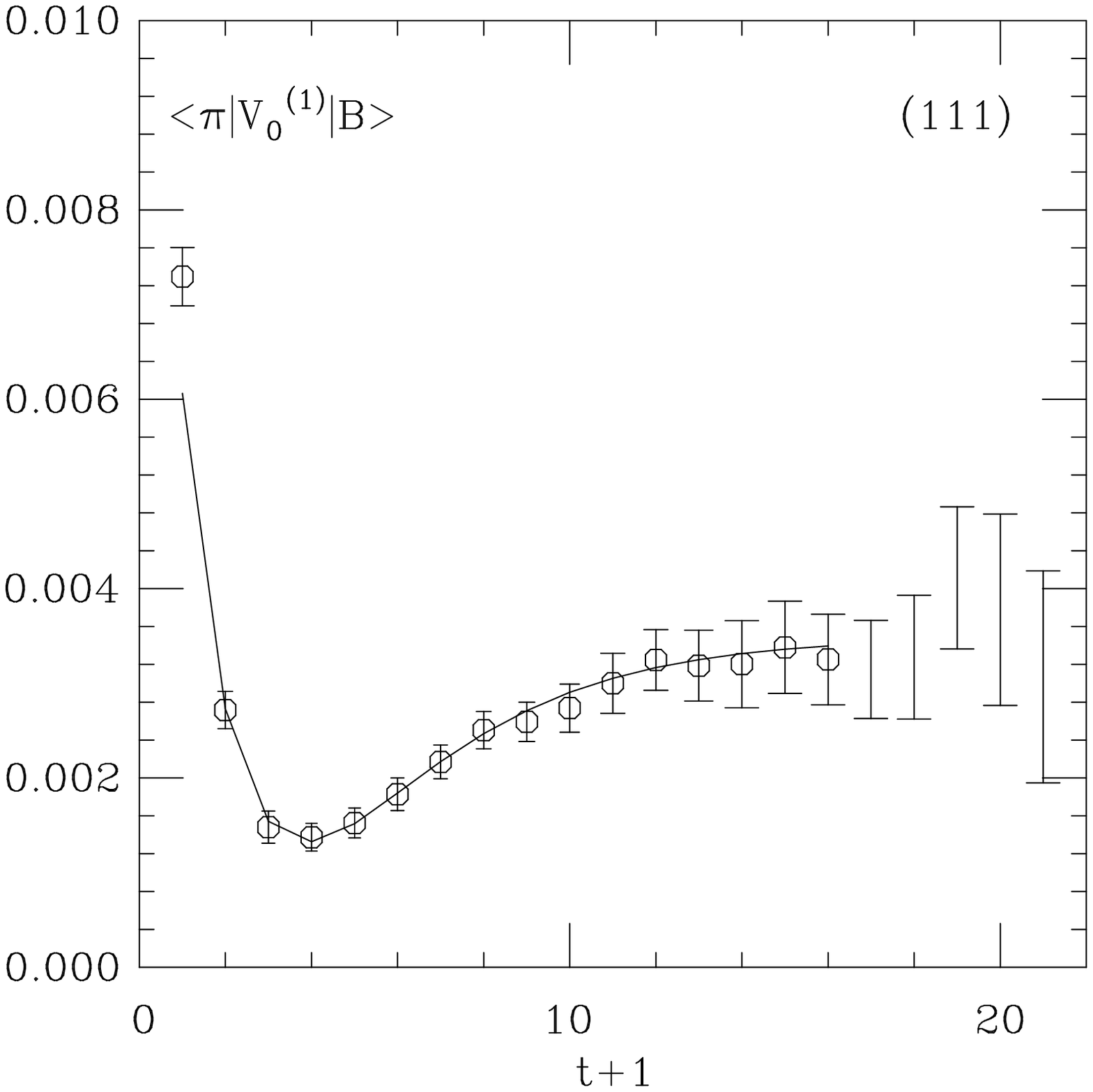}{90mm}
}
\caption{same as Fig.4 for $V_0^{(1)}$, the tree-level 1/M current
 correction }
\end{figure}

\begin{figure}
\begin{center}
\epsfysize=7.in
\centerline{\epsfbox{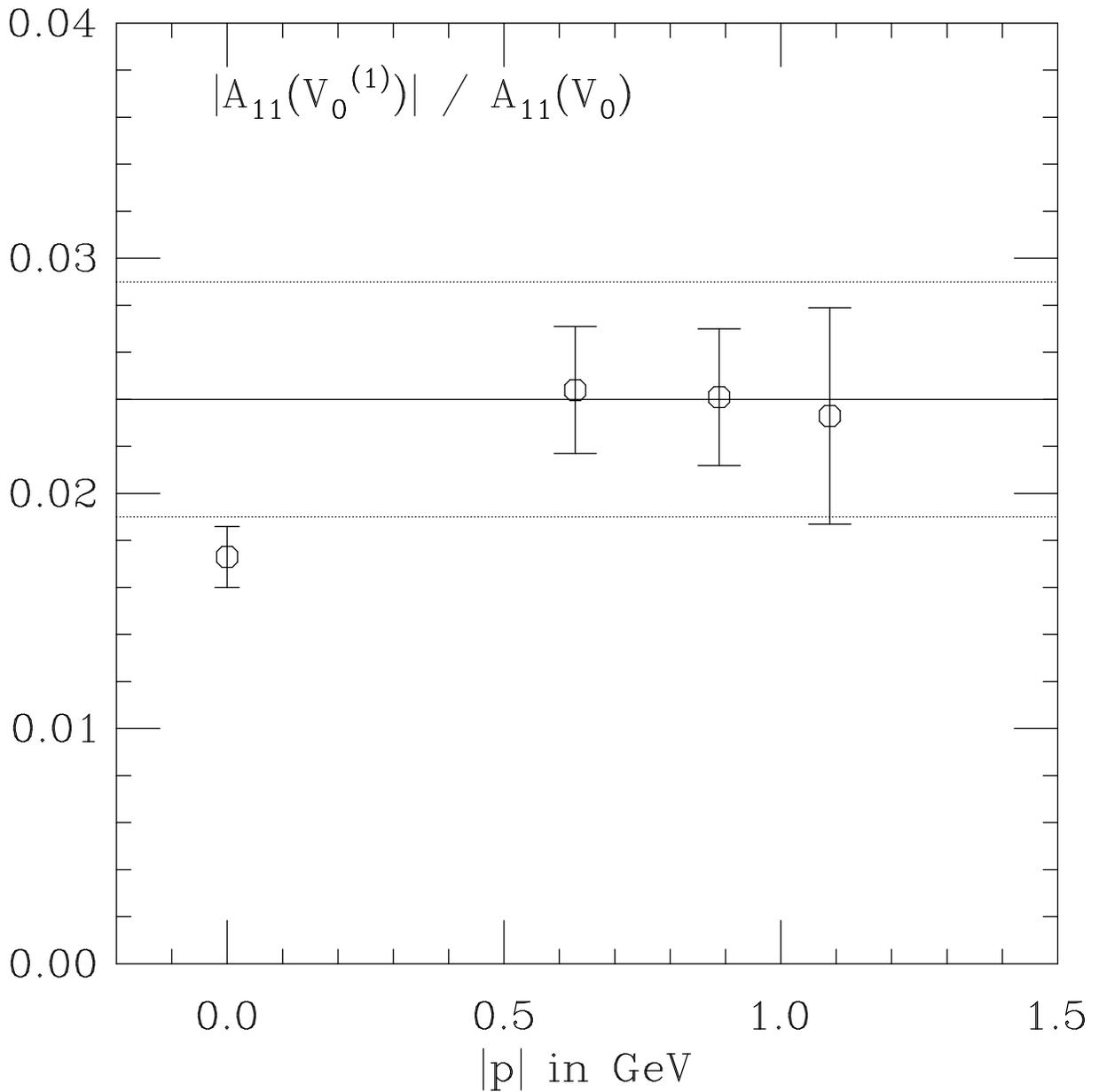}}
\end{center}
\caption{Ratio of groundstate contributions to $\langle V^{(1),L}_0 
\rangle$ and $\langle V^L_0 \rangle$ for several pion momenta. The 
horizontal line shows the one-loop $O(\alpha_s/a_sM)$ power law subtraction 
term for the $\langle V^{(1),L}_0 \rangle$ matrix element.
  }
\end{figure}

\begin{figure}
\begin{center}
\epsfysize=7.in
\centerline{\epsfbox{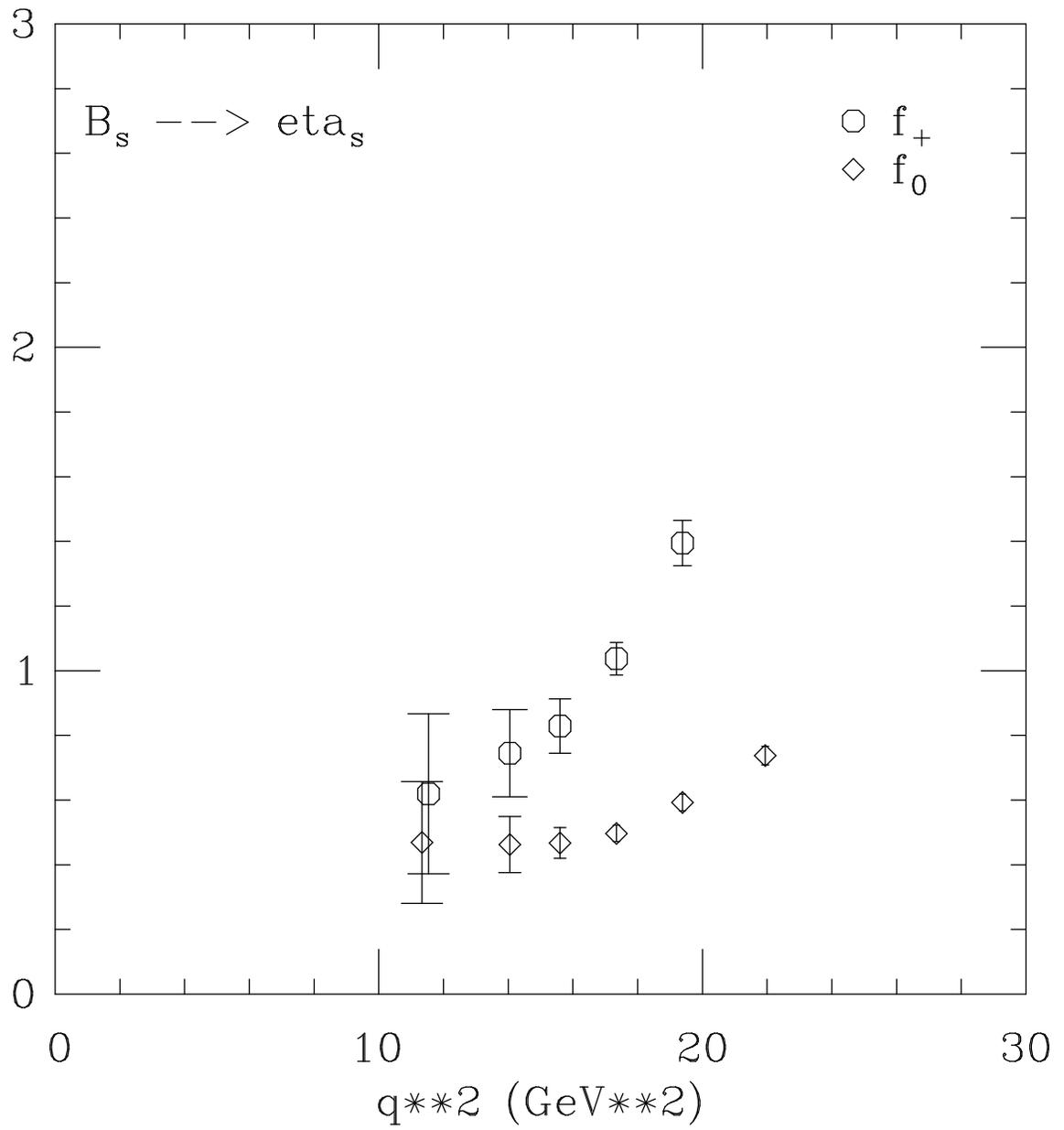}}
\end{center}
\caption{The form factors $f_+$ and $f_0$ for the light quark mass fixed 
at the strange quark mass.  Only statistical errors are shown.
  }
\end{figure}

\begin{figure}
\begin{center}
\epsfysize=7.in
\centerline{\epsfbox{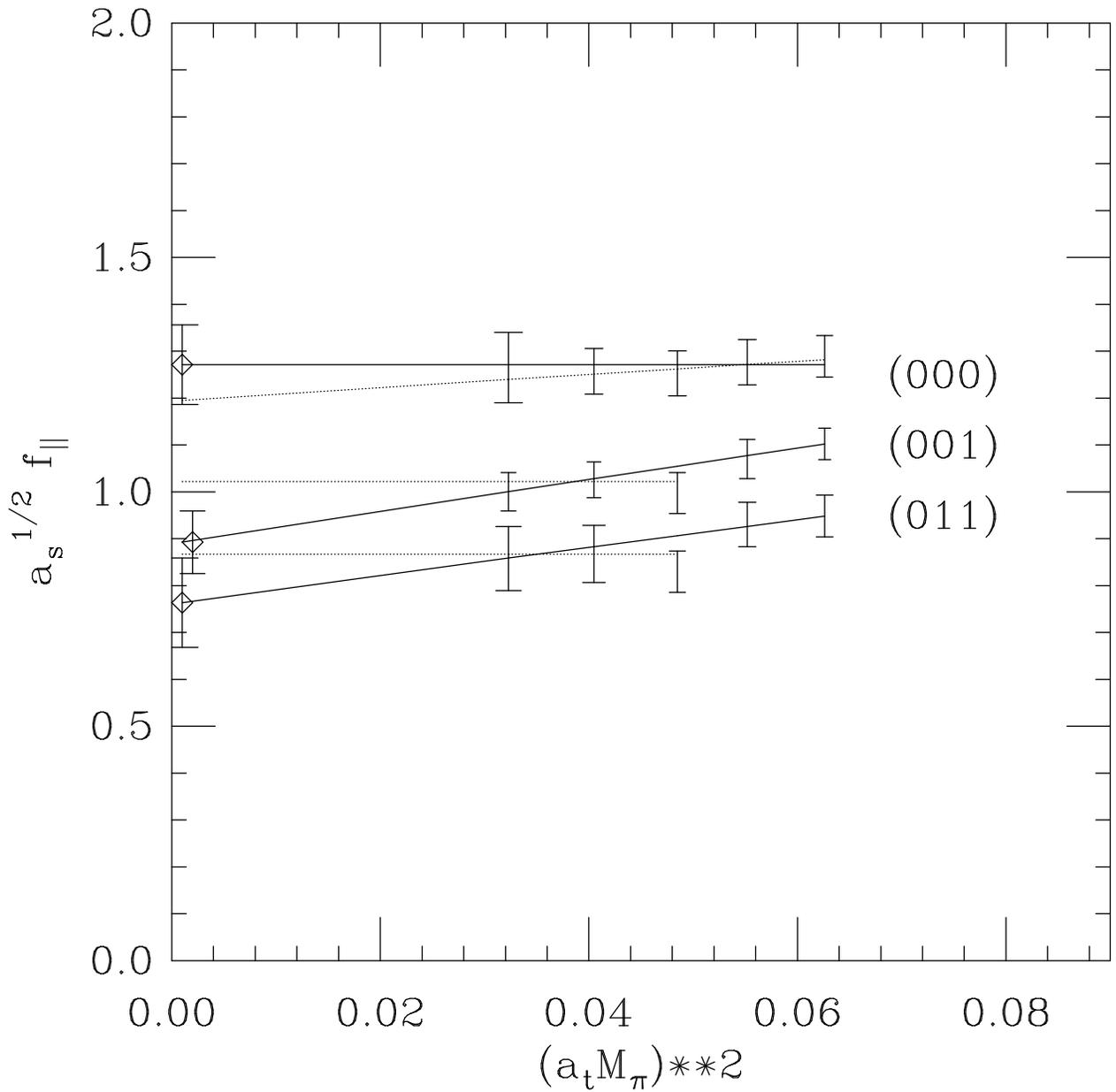}}
\end{center}
\caption{ Chiral extrapolations of the form factor $f_\|$ at fixed 
pion momentum. Constant and linear fits were carried out to all 5 
or to the last 3 data points.  The full and dotted lines give some 
idea of spread in fit results.
  }
\end{figure}

\begin{figure}
\begin{center}
\epsfysize=7.in
\centerline{\epsfbox{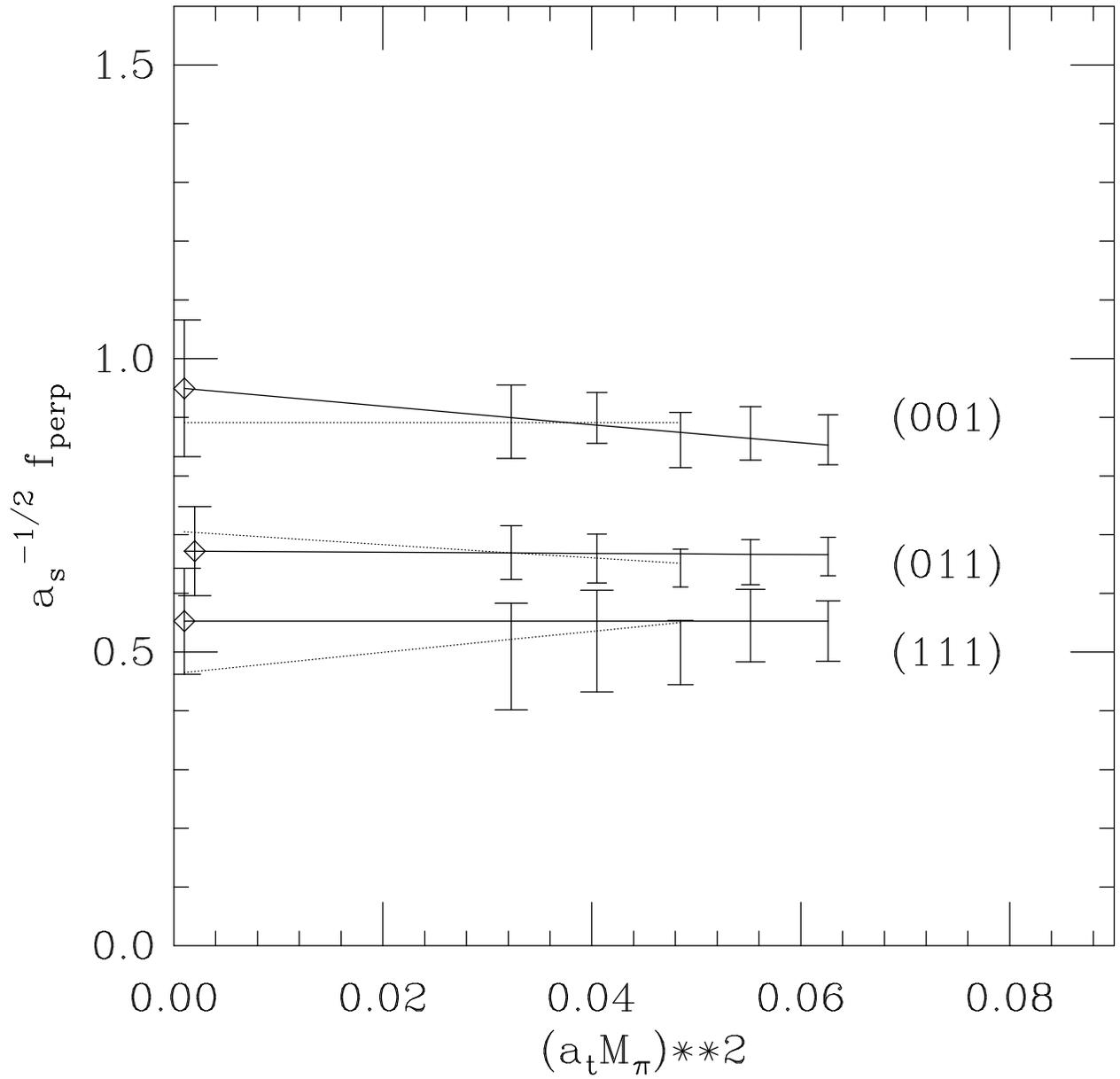}}
\end{center}
\caption{ Same as Fig.13 for the form factor $f_\bot$.
  }
\end{figure}

\begin{figure}
\begin{center}
\epsfysize=7.in
\centerline{\epsfbox{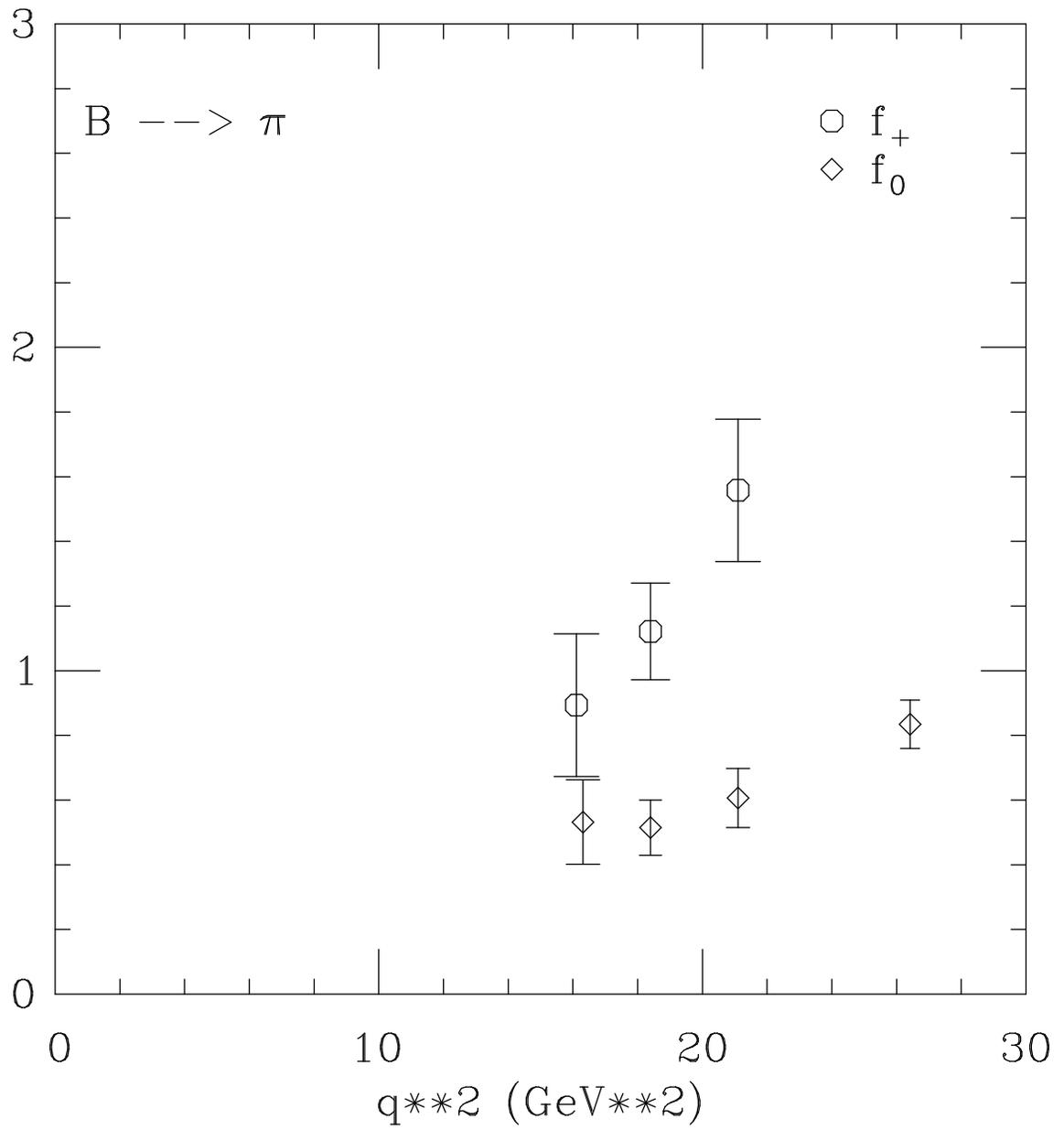}}
\end{center}
\caption{The form factors $f_+$ and $f_0$ after chiral extrapolation 
to the physical pion.
  Statistical and chiral extrapolation errors shown.
  }
\end{figure}

\begin{figure}
\begin{center}
\epsfysize=7.in
\centerline{\epsfbox{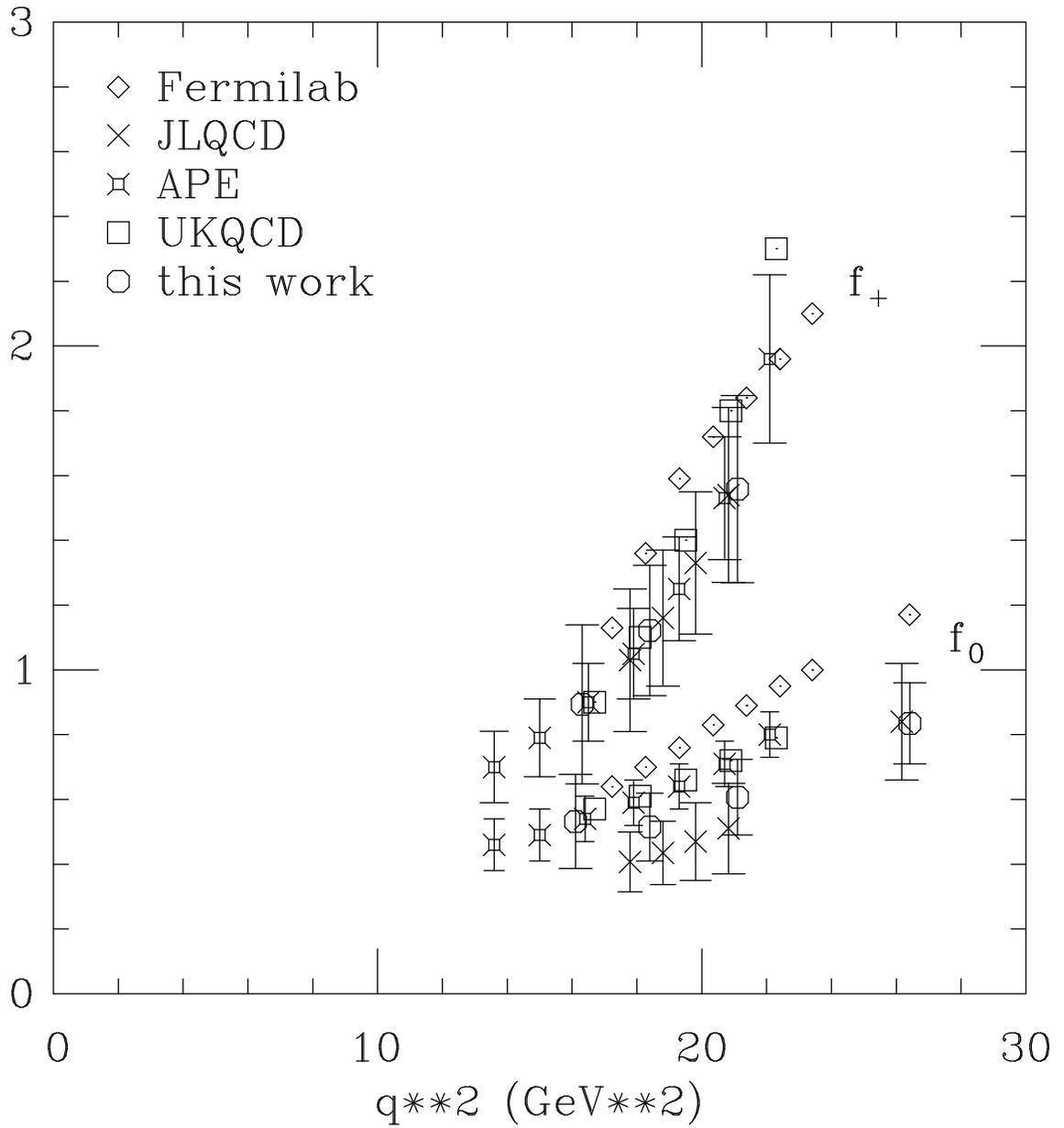}}
\end{center}
\caption{Comparison with other groups. 
  Statistical, chiral extrapolation and other systematic errors included.
To avoid too much clutter we do not include errors for the Fermilab and 
UKQCD data points.  They are comparable to those of other groups.
  }
\end{figure}

\begin{figure}
\begin{center}
\epsfysize=7.in
\centerline{\epsfbox{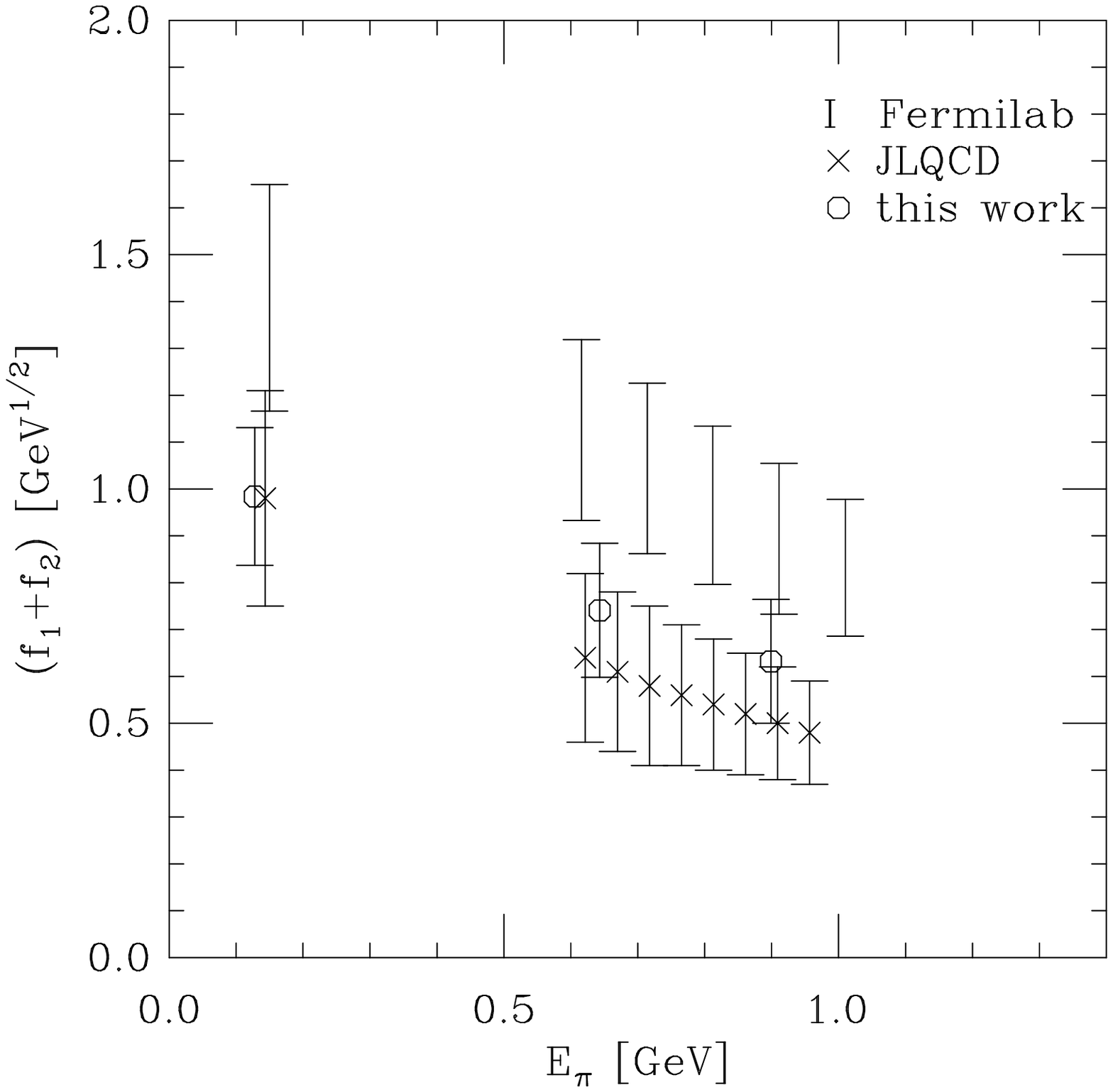}}
\end{center}
\caption{Comparison with the Fermilab and JLQCD collaborations
for the form factor $(f_1+f_2)$.
  }
\end{figure}

\begin{figure}
\begin{center}
\epsfysize=7.in
\centerline{\epsfbox{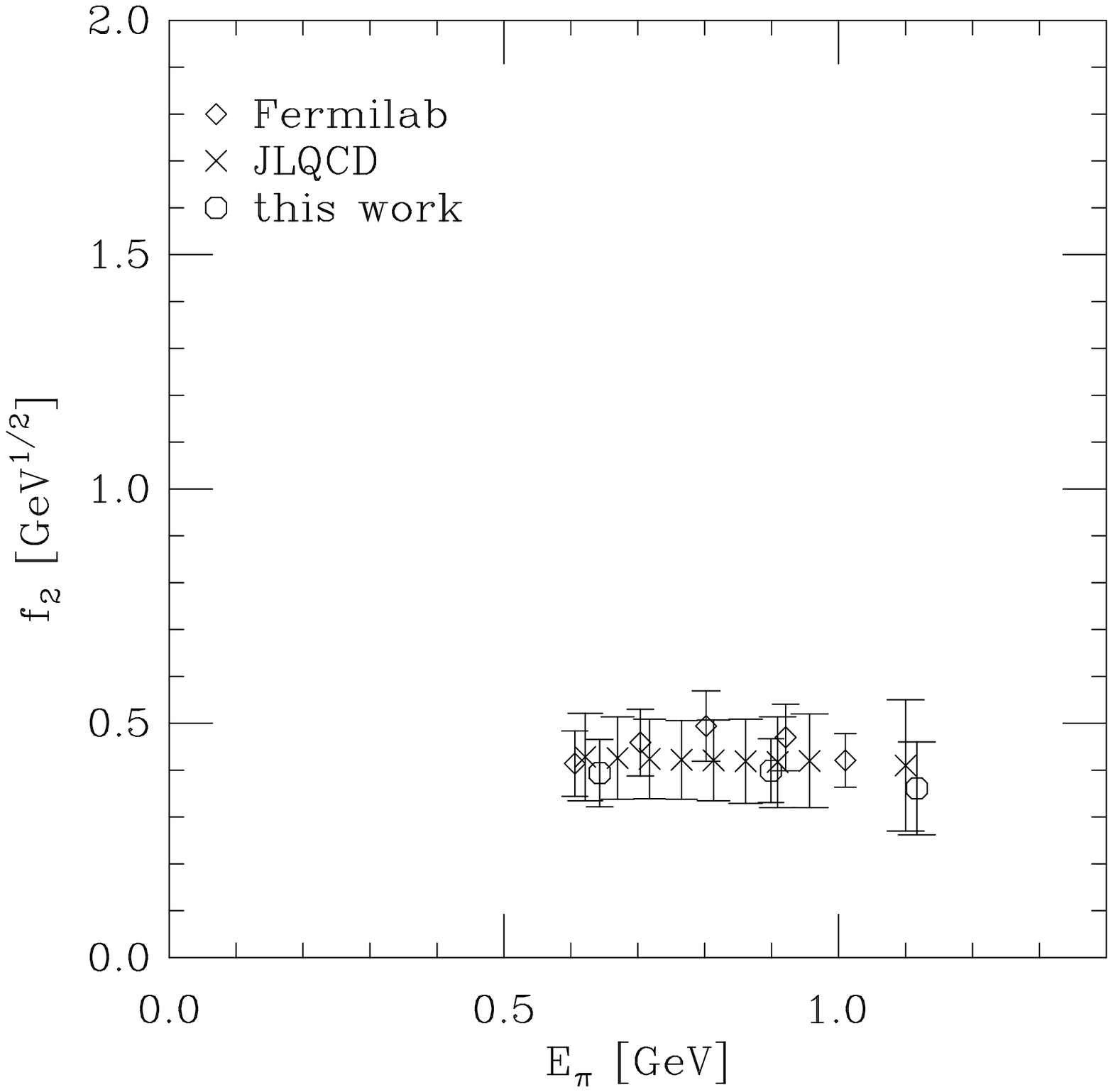}}
\end{center}
\caption{Comparison with the Fermilab and JLQCD collaborations
for the form factor $f_2$.
  }
\end{figure}

\begin{figure}
\centerline{
\ewxy{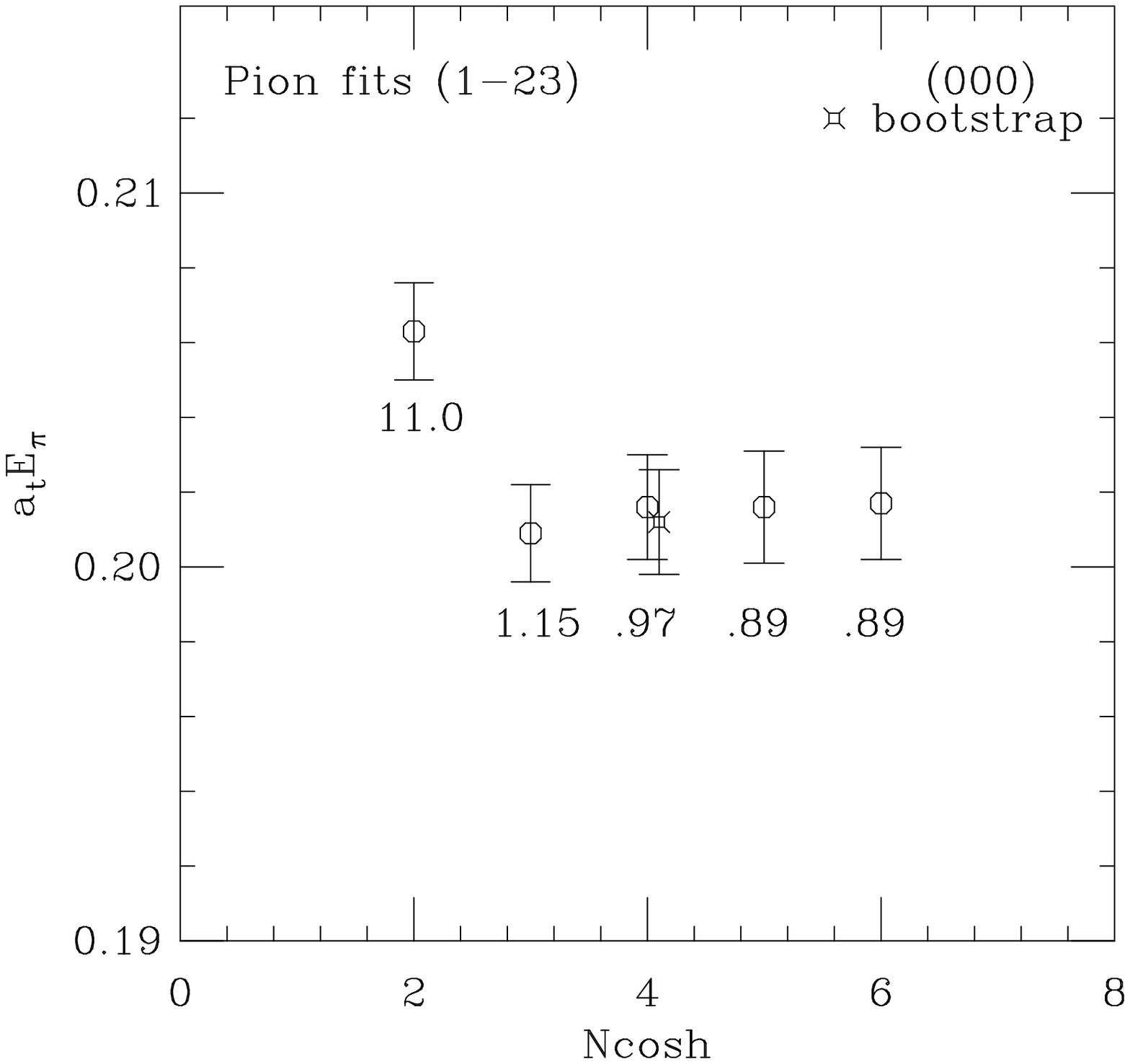}{90mm}
\ewxy{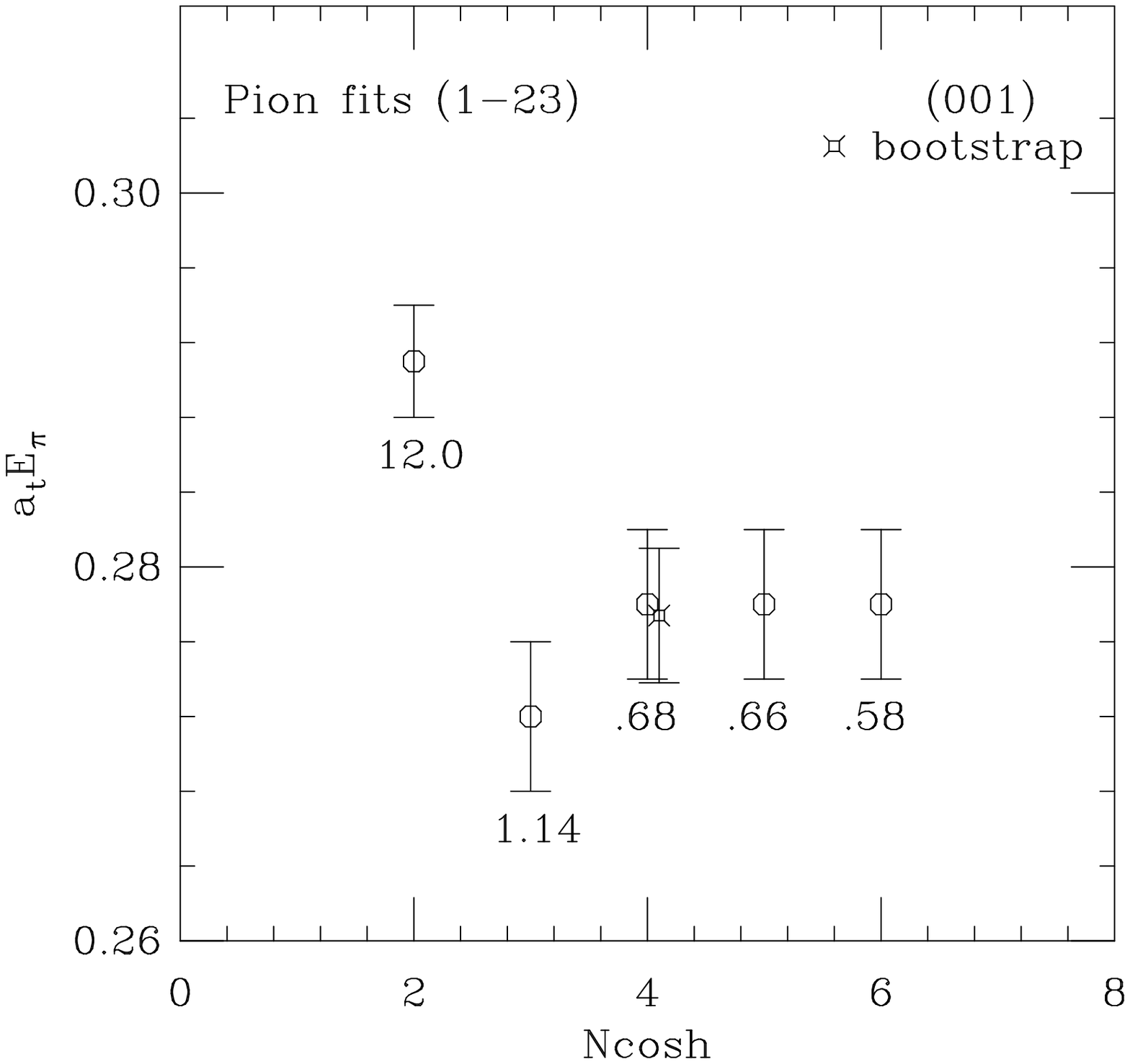}{90mm}
}
\caption{Fit results for groundstate pion energies versus $N_{cosh}$.
Pion momentum is shown on upper right and the fit range on  upper left 
corners. The fancy star shows bootstrap fit results. The numbers below 
data points give $\chi^2/d.o.f$. 
 }
\end{figure}

\begin{figure}
\centerline{
\ewxy{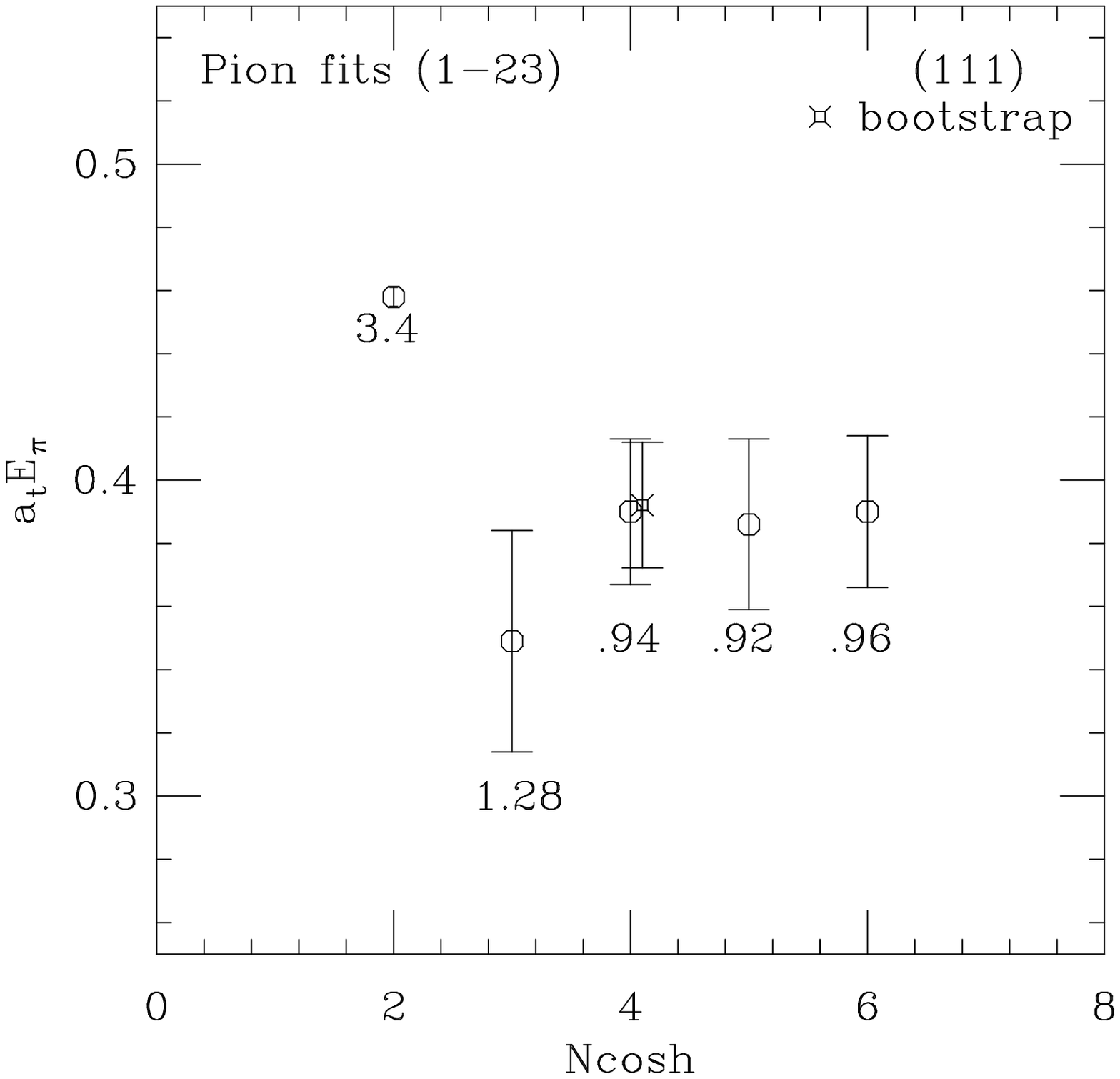}{90mm}
\ewxy{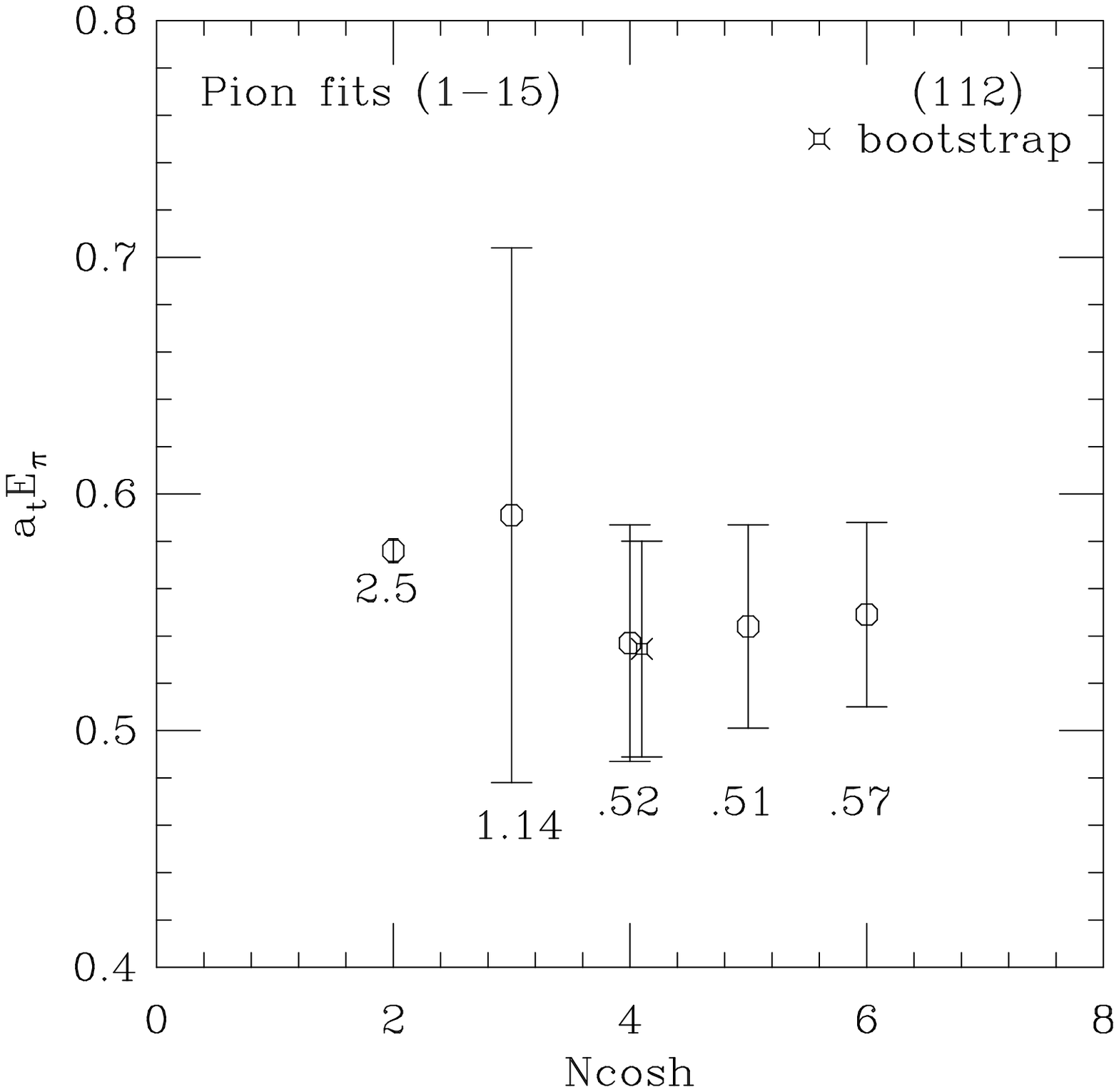}{90mm}
}
\caption{Same as Fig.19 for pion momenta (111) and (112).
 }
\end{figure}

\begin{figure}
\centerline{
\ewxy{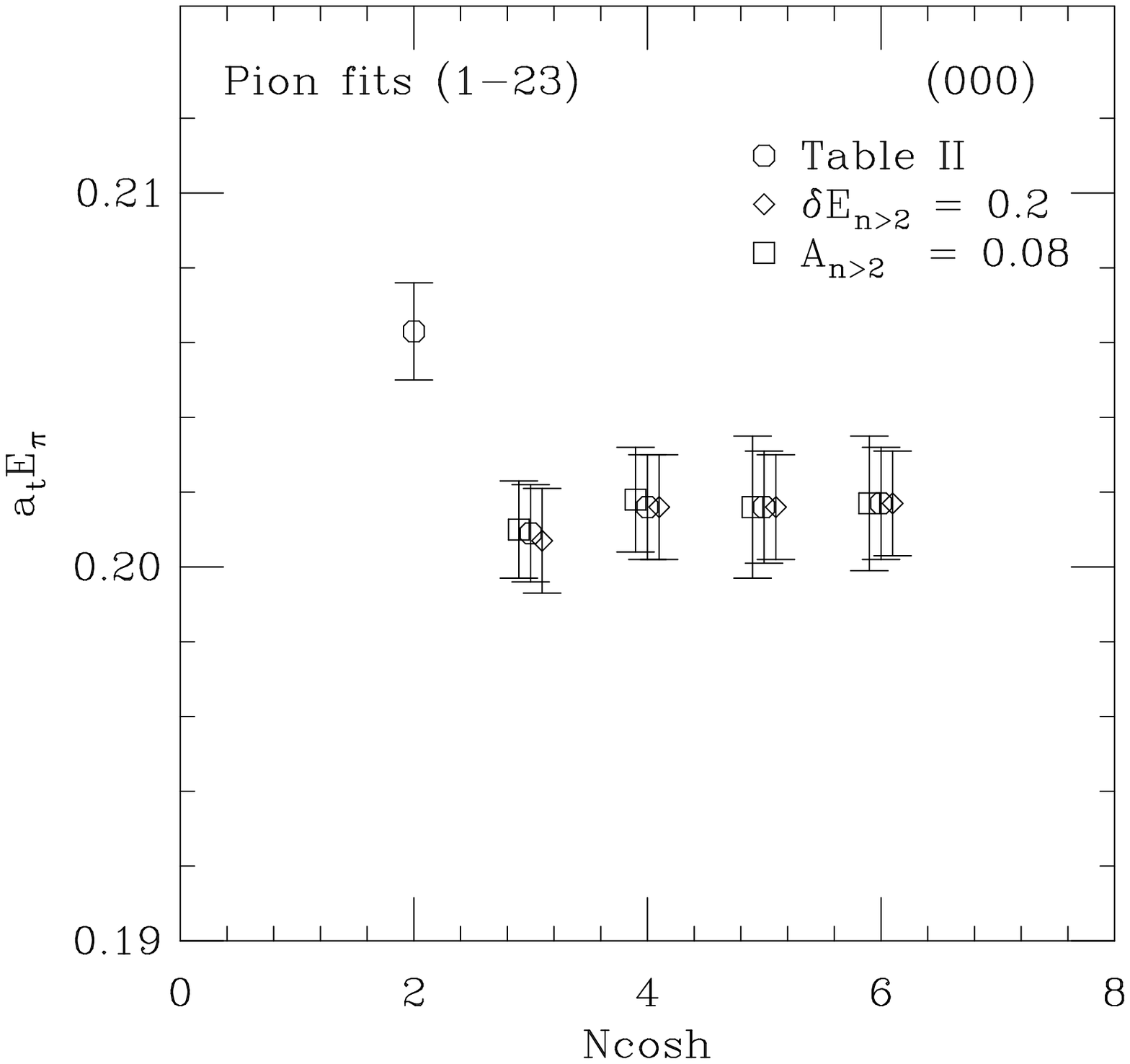}{90mm}
\ewxy{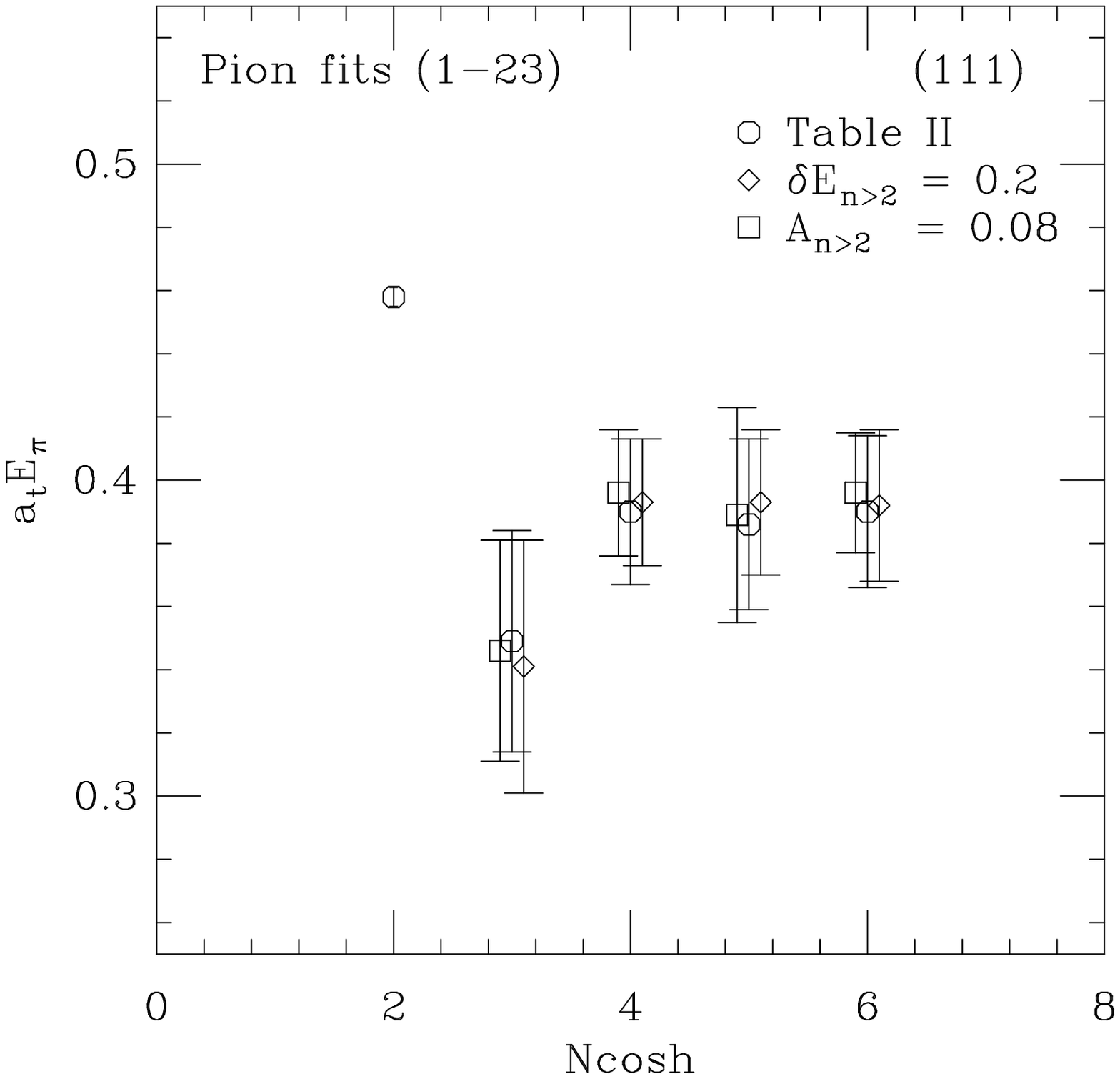}{90mm}
}
\caption{Comparisons of fits using priors of Table II with fits 
after changing the $n>2$ priors as indicated.
 }
\end{figure}

\begin{figure}
\centerline{
\ewxy{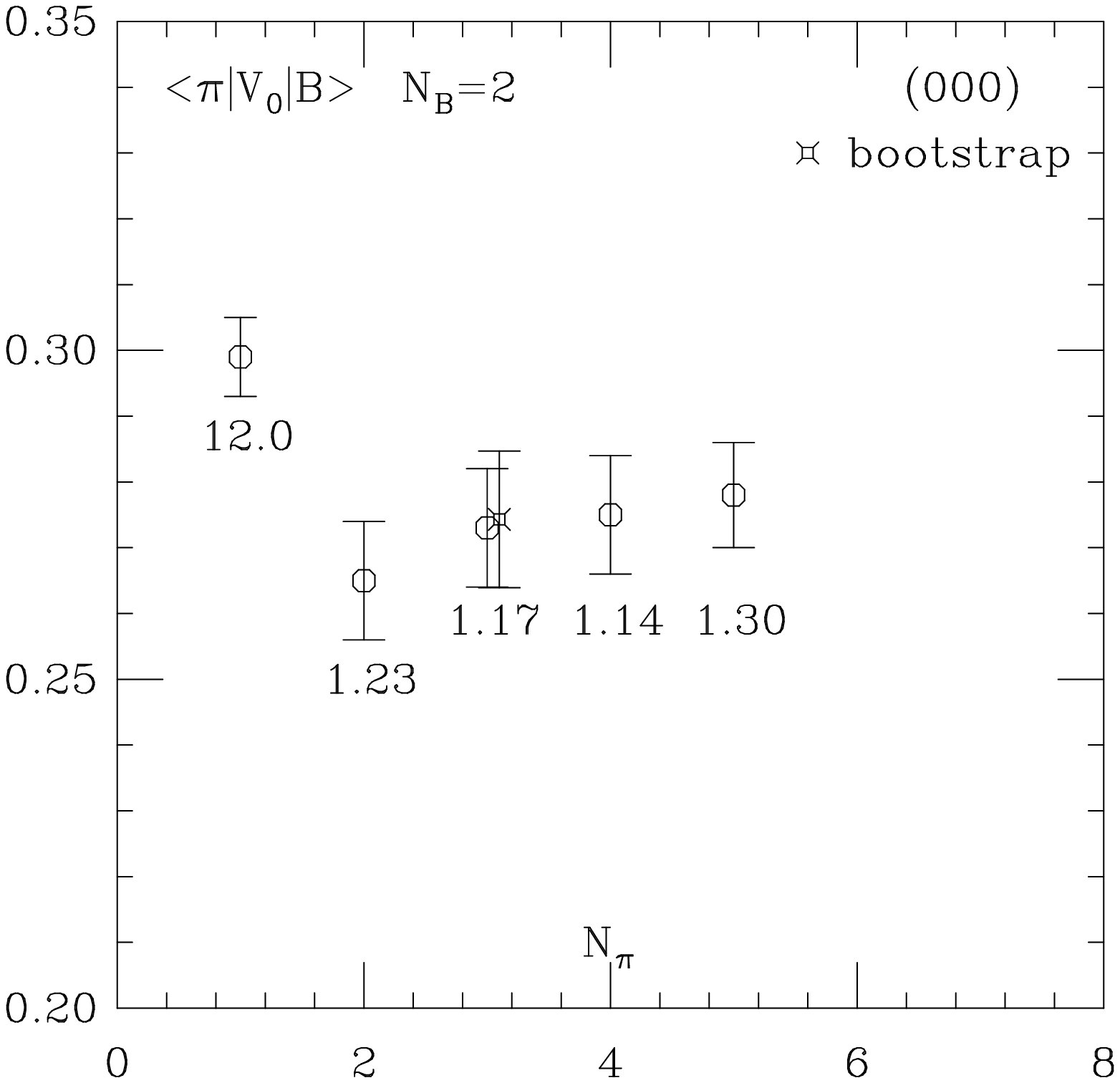}{90mm}
\ewxy{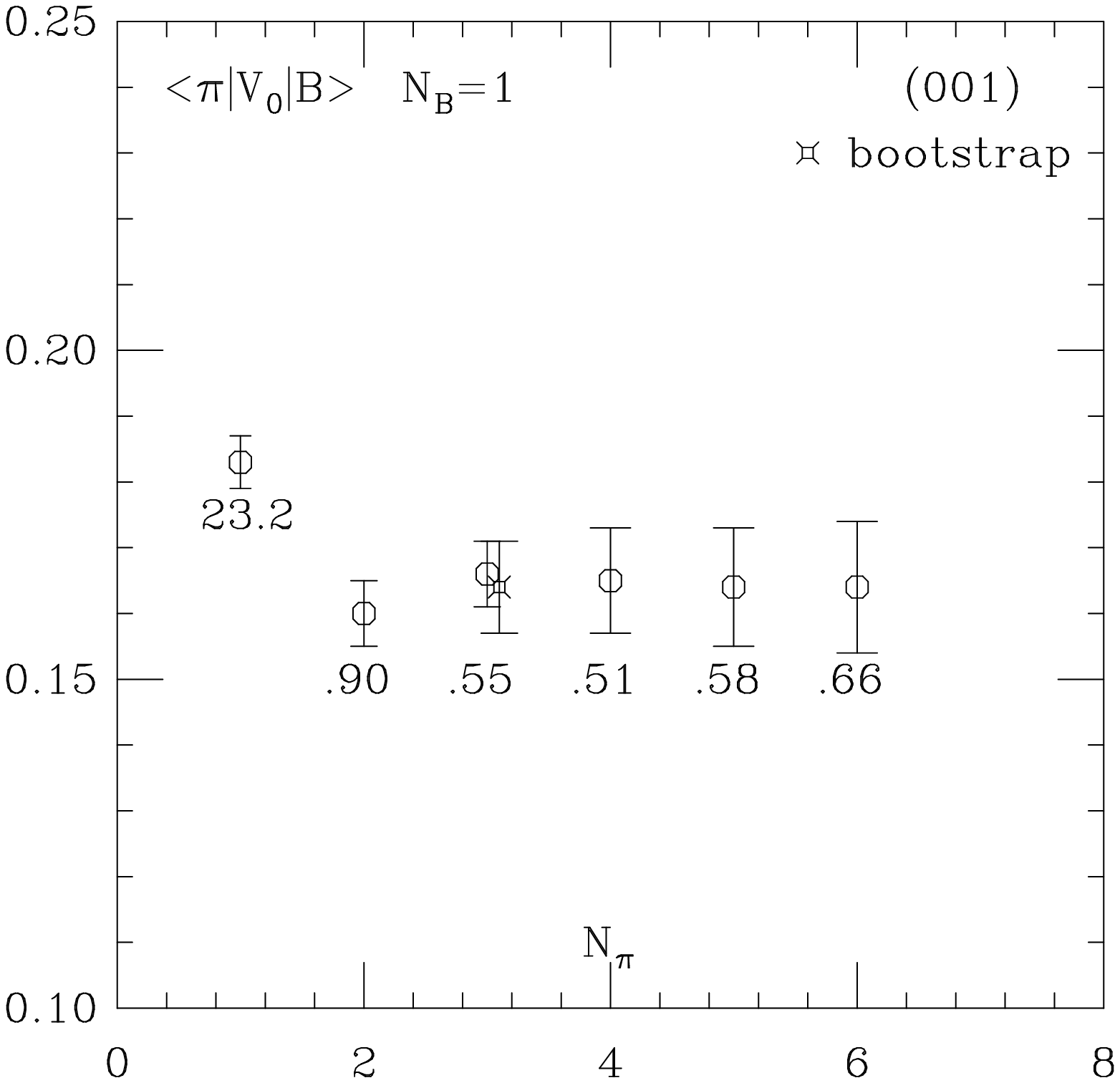}{90mm}
}
\caption{Fit results for the groundstate amplitude $A_{11}$ from 
the $\langle V_0 \rangle$ threepoint correlator. The numbers below 
the data points give $\chi^2/d.o.f$.
 }
\end{figure}

\begin{figure}
\centerline{
\ewxy{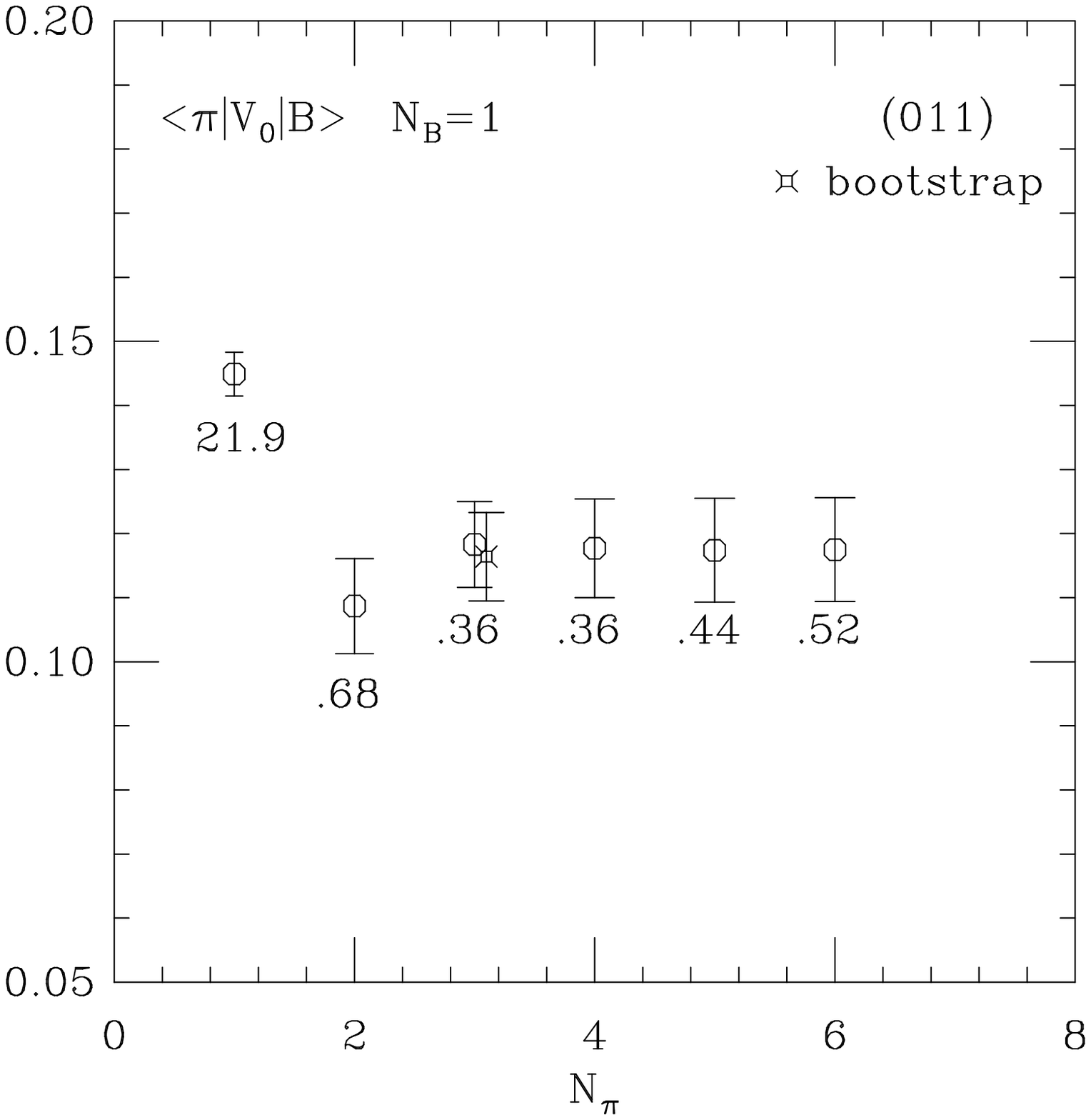}{90mm}
\ewxy{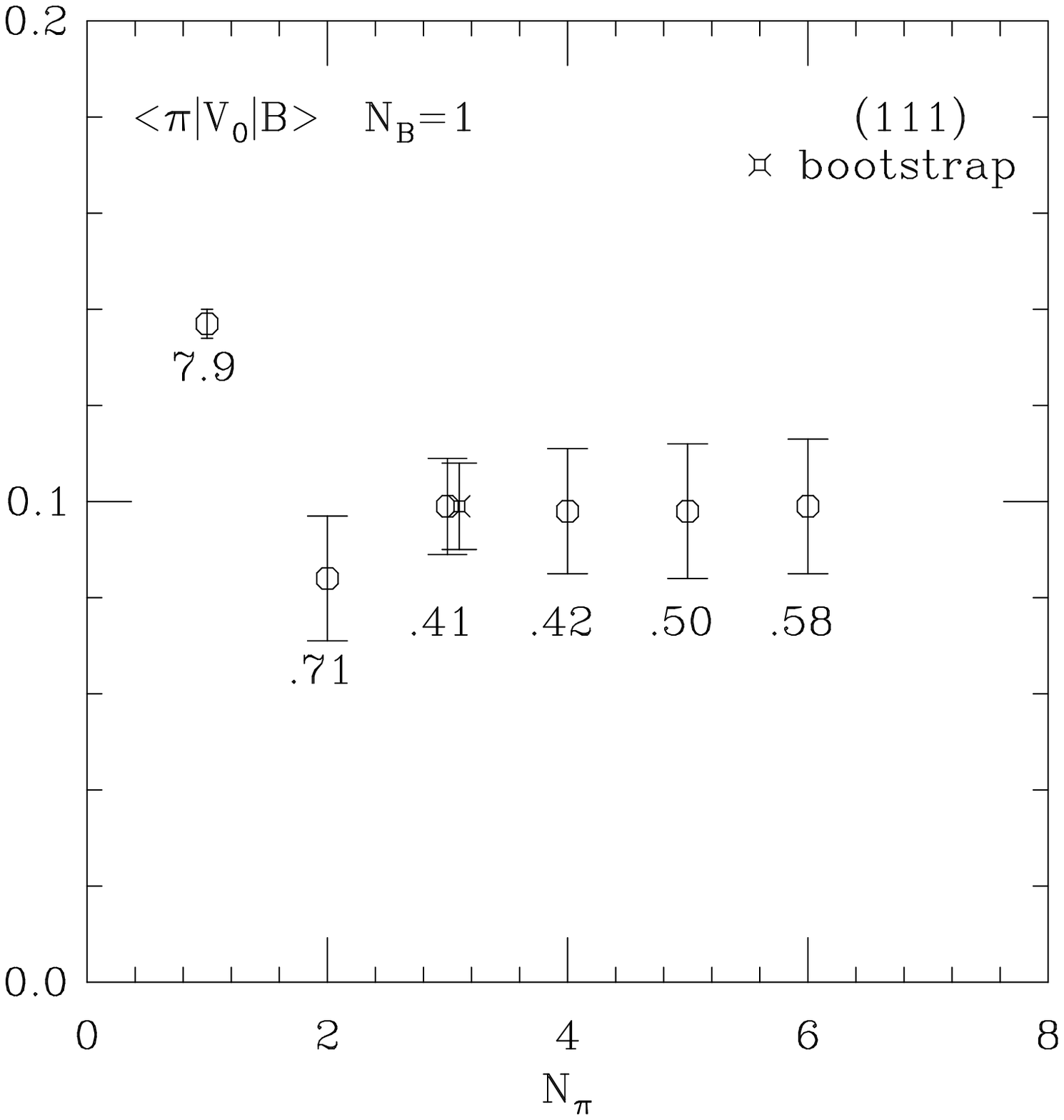}{90mm}
}
\caption{Same as Fig.22 for momenta (011) and (111).
 }
\end{figure}


\begin{thebibliography}{8}

\bibitem{fnal}
A. El-Khadra {\em et al.} [Fermilab Collaboration],
 Phys. Rev. {\bf D64}:014502 (2001).

\bibitem{jlqcd}
S. Aoki {\em et al.} [JLQCD Collaboration], 
 Phys. Rev. {\bf D64}:114505 (2001).

\bibitem{ape}
A. Abada {\em et al.} [APE Collaboration] , Nucl. Phys. {\bf B619}, 565 
(2001).

\bibitem{ukqcd}
K.C. Bowler {\em et al.} [UKQCD Collaboration], Phys. Lett. {\bf B486} 
(2000) 111.

\bibitem{mikecolin}
  C.~Morningstar and M.~Peardon, Phys.\ Rev.{\bf D60}:034509 (1999).

\bibitem{manke}
 T.~Manke {\em et al.}, Phys. \ Rev. \ Lett {\bf 82}:4396 (1999).

\bibitem{ianron}
  I.~Drummond {\em et al.},
Phys. Lett. {\bf B478}, 151 (2000).

\bibitem{comparison}
S. Collins {\em et al.}, 
 Phys. Rev. {\bf D64}:055002 (2001).

\bibitem{bayes}
G.P. Lepage {\em et al.}, 
 Nucl. Phys. {\bf B}(Proc. \ Suppl.){\bf 106}, 12 (2002).

\bibitem{staggered}
S. Naik, 
 Nucl. \ Phys. {\bf B316}, 238 (1989).  \\
G. P. Lepage, Phys. Rev. {\bf D59}:074501 (1999). \\
K. Orginos, D. Toussaint and R.L. Sugar, Phys. Rev. {\bf D60}:054503 (1999).

\bibitem{matt}
M. Wingate {\em et al.},
 Nucl. Phys. {\bf B}(Proc. \ Suppl.){\bf 106}, 379 (2002);  \\
M. Wingate; talk presented at ``Lattice 02''.

\bibitem{mnrqcd}
K. Foley;  poster presented at ``Lattice 02''.


\bibitem{markron}
M. Alford {\em et al.}
 Phys. Rev. {\bf D63}:074501 (2001).

\bibitem{alford}
M.~Alford, T.~Klassen and G.~P.~Lepage, Phys. \ Rev. \ {\bf D58}:034503 
(1998), Nucl. \ Phys. {\bf B496} 377 (1997).


\bibitem{pert}
S. Groote and J. Shigemitsu, Phys. Rev. {\bf D62}:014508 (2000).


\bibitem {cornell}
G.P. Lepage {\em et al.}, 
 Phys. Rev. {\bf D46}, 4052 (1992).

\bibitem{ianron2}
  I.~Drummond {\em et al.},
 Nucl. Phys. {\bf B}(Proc. \ Suppl.){\bf 73}, 336 (1999).

\bibitem{hartron}
A. Hart and R.R. Horgan; poster presented at ``Lattice 02''.

\bibitem{martin}
M. Guertler {\em et al.};
 Nucl. Phys. {\bf B}(Proc. \ Suppl.){\bf 106}, 409 (2002).

 \bibitem {pert2}
C. Morningstar and J. Shigemitsu;  Phys. Rev. {\bf D57}, 6741 (1998);
  Phys. Rev. {\bf D59}:094504 (1999).

\bibitem{fbscale}
S. Collins {\em et al.}, 
 Phys. Rev. {\bf D63}:034505 (2001).

\bibitem{claude}
C. Bernard; talk presented at ``Lattice 02''.

\end{thebibliography}
\end{document}